\documentclass[a4paper,fleqn,usenatbib]{mnras}
\usepackage[dvipdfmx]{graphicx}
\usepackage{newtxtext,newtxmath}

\usepackage[T1]{fontenc}
\usepackage{ae,aecompl}

\usepackage{color}
\usepackage[dvipsnames]{xcolor}
\usepackage{ulem}

\usepackage{graphicx}	
\usepackage{amsmath}	
\usepackage{amssymb}	

\graphicspath{{./figure/}}

\newcommand{\dtog}{\mathcal{D}}   
\newcommand{\stol}{\mathcal{D}_\mathrm{S} / \mathcal{D}_\mathrm{L}}
\newcommand{\RNum}[1]{\uppercase\expandafter{\romannumeral #1\relax}}
\newcommand{\Msun}{{{\rm M}_\odot}}

\title[Dust in cosmological simulation]{Dust scaling relations in a cosmological simulation}

\author[K. C. Hou et al.]
{Kuan-Chou Hou$^{1,2}$\thanks{E-mail: hou@post.bgu.ac.il},
Shohei Aoyama$^{1}$,
Hiroyuki Hirashita$^{1}$,
Kentaro Nagamine$^{3,4,5}$,
\newauthor and
Ikkoh Shimizu$^{3}$\\
$^{1}$Institute of Astronomy and Astrophysics, Academia Sinica,
PO Box 23-141, Taipei 10617, Taiwan\\
$^{2}$Physics Department, Ben-Gurion University of the Negev,
Be'er-Sheva 84105, Israel\\
$^{3}$Theoretical Astrophysics, Department of Earth \& Space Science,
Osaka University, 1-1 Machikaneyama, Toyonaka, Osaka 560-0043, Japan\\
$^{4}$Department of Physics \& Astronomy, University of Nevada Las Vegas,
4505 S. Maryland Pkwy, Las Vegas, NV 89154-4002, USA \\
$^{5}$Kavli IPMU (WPI), University of Tokyo, 5-1-5 Kashiwanoha, Kashiwa, Chiba, 277-8583, Japan \\
}

\date{Accepted XXX. Received YYY; in original form ZZZ}

\pubyear{2018}

\begin{document}
\label{firstpage}
\pagerange{\pageref{firstpage}--\pageref{lastpage}}
\maketitle

\begin{abstract}

To study the dust evolution in the cosmological structure formation history,
we perform a smoothed particle hydrodynamic simulation
with a dust enrichment model in a cosmological volume.
We adopt the dust evolution model
that represents the grain size distribution by two sizes
and takes into account stellar dust production and
interstellar dust processing.
We examine the dust mass function and the scaling properties of dust
in terms of the characteristics of galaxies.
The simulation broadly reproduces the observed dust mass functions
at redshift $z = 0$,
except that it overproduces the massive end
at dust mass $M_\mathrm{d} \gtrsim 10^{8}$\,$\Msun$.
This overabundance is due to overproducing massive
gas/metal-rich systems, but we also note that the relation between
stellar mass and gas-phase metallicity
is reproduced fairly well by our recipe.
The relation between dust-to-gas ratio and metallicity shows
a good agreement with the observed one at $z=0$,
which indicates successful
implementation of dust evolution in our cosmological simulation.
Star formation consumes not only gas but also dust,
causing a decreasing trend of the dust-to-stellar mass ratio
at the high-mass end of galaxies.
We also examine the redshift evolution up to $z \sim~ 5$,
and find that the galaxies have on average the highest dust mass at $z = 1-2$.
For the grain size distribution,
we find that galaxies with metallicity $\sim 0.3~ Z_\odot$
tend to have the highest small-to-large grain abundance ratio;
consequently, the extinction curves in those galaxies have the steepest
ultraviolet slopes.

\end{abstract}




\section{Introduction}
\label{sec:Intro}

Dust has been observed in both local and high-redshift galaxies,
and is known to be important in understanding galaxy evolution.
Since dust absorbs the ultraviolet (UV) light from stars
and reemits it in the far-infrared (FIR),
the spectral energy distribution (SED) of galaxies is strongly modified by dust
\citep[e.g.][for recent modelling]{Yajima:2014aa,Schaerer:2015aa}.
Thus, the star formation rate (SFR) derived from stellar UV emission must
be corrected for dust extinction, and the SFR estimated at FIR wavelengths
is complementary in tracing the obscured star formation activities
\citep{Buat:1996aa,Hirashita:2003aa}.
Moreover, dust surfaces are the main sites for efficiently producing
molecular hydrogen (H$_2$), which is the dominant constituent of
molecular clouds and an important coolant in low-metallicity
environments \citep{Hirashita:2002aa,Cazaux:2004aa}.
Moreover, the characteristic mass of the final star-forming fragments
is determined by dust cooling \citep{Omukai:2005aa,Schneider:2006aa}.
Therefore, star formation activities and their emission properties
in galaxies are strongly affected by dust.

Studies of the cosmic background radiation
showed that
the total radiation energy in the FIR is comparable
to that in the optical \citep[e.g.][]{Hauser:1998aa}.
This means that reprocessed stellar light by dust is important
in tracing the total radiation energy emitted by stars.
Many studies have reported that the cosmic star formation activity
is more obscured at redshifts $z\sim 1-2$ compared to the local Universe
\citep{Hopkins:2001aa,Sullivan:2001aa,Takeuchi:2005aa,Burgarella:2013aa}.
In the past decades, many submillimetre galaxies,
which have extremely high SFRs and a
significant amount of dust, have been observed in the distant Universe
\citep{Blain:2002aa,Casey:2014aa}.
The Atacama Large Millimeter/submillimeter Array (ALMA) has
detected dust emission from `normal' galaxies
at redshift $z > 6$
\citep{Watson:2015aa,Willott:2015aa,Laporte:2017aa,Hashimoto:2018aa,Tamura:2018aa}.
However, dust continuum emission has not been detected by ALMA
for a large fraction of galaxies at $z \gtrsim 6$
\citep{Ouchi:2013aa,Schaerer:2015aa,Bouwens:2016aa,
Inoue:2016aa,Carniani:2018aa}, which implies that
there is a large variety in the dust abundance among high-redshift galaxies.

Dust properties in galaxies are usually studied
by scaling relations regarding dust-to-gas ratio
\citep{Lisenfeld:1998aa,Draine:2007ab,Galametz:2011aa,Remy-Ruyer:2014aa}
and dust-to-stellar mass ratio
\citep{Dunne:2011aa,Clark:2015aa,Calura:2017aa}.
In particular, the relation between dust-to-gas ratio and metallicity
is often used to investigate the dust evolution processes
or to test theoretical dust evolution models, since metallicity is an indicator of
chemical enrichment driving the dust evolution \citep{Inoue:2011aa}.
\citet{Remy-Ruyer:2014aa} studied the local galaxies covering
a wide metallicity range, and found that the relation between
dust-to-gas ratio and metallicity is not
represented well by a single power-law, but is better explained
by a double power-law with a break around $Z \sim 0.1~ Z_\odot$.
They suggested that not only dust production in stellar ejecta
[supernovae (SNe) and asymptotic giant branch (AGB) star winds]
but also dust growth by accretion in the interstellar medium (ISM)
is required in describing the relation.
Many authors revealed that dust produced in stellar ejecta
is not enough to explain the total dust abundance in galaxies;
thus, dust growth in the ISM is claimed to
play an important role in both the local
\citep{Dwek:1998aa,Zhukovska:2008aa,Hirashita:2011aa,Asano:2013ab}
and distant Universe
\citep{Valiante:2011aa,Mancini:2015aa,Michaowski:2015aa,Nozawa:2015aa,Popping:2017aa}.

The grain size distribution is also a fundamental property of dust
\citep{Mathis:1977aa,Liffman:1989aa},
since it directly influences some dust-related observational quantities
such as extinction curves
\citep{Draine:1984aa,Weingartner:2001aa,Hirashita:2009aa,Asano:2014aa,Hou:2016aa,Hou:2017aa}
and affects molecular formation \citep{Yamasawa:2011aa,Harada:2017aa,Chen:2018aa},
and dust evolution \citep{Kuo:2012aa,Asano:2013aa,Hirashita:2015ab}.

There have been some efforts of modelling the evolution of grain size distribution
in the ISM.
\citet{Hirashita:2009aa} investigated shattering and coagulation in turbulent ISM.
These processes affect not only the grain size distribution
but also the total dust abundance because the surface-to-volume ratio
changes the SN destruction rate and grain growth rate
\citep[e.g.][]{Hirashita:2011aa}.
\citet{Asano:2013aa} established a full framework
for calculating the grain size distribution in
a consistent manner with chemical enrichment over
the entire galaxy history.
\citet{Asano:2014aa} used their model to calculate
the evolution of extinction curves, producing steeper extinction curves
than that of the Milky Way at an appropriate age for the current Universe
($\sim 10$ Gyr).
\citet{Nozawa:2015aa} successfully reproduced the Milky Way
extinction curve using the same framework but including a dense
cloud component which hosts efficient coagulation.
This indicates that dust evolution is sensitive
to the physical condition of the ISM.
Moreover, the variation of the Milky Way extinction curves
toward different lines of sight indicates inhomogeneity
of dust properties in the galaxy \citep{Fitzpatrick:2007aa}.
Although the above dust evolution models take a great step forward to
the understanding of dust evolution, they treated a galaxy as
a single zone and neglected the spatial inhomogeneity in the ISM.
Thus, the spatial diversity of dust properties
is still a challenge in the current frontier of dust evolution modelling.

Hydrodynamic simulations have been providing useful insight into
galaxy formation and evolution.
Since hydrodynamic simulations solves gas dynamics and
computes the physical conditions in galaxies,
they are useful to study dust formation and evolution in
a consistent manner with the physical state of the ISM.
\citet{Yajima:2015aa} performed zoom-in cosmological hydrodynamic
simulations of galaxies with a fixed dust-to-metal ratio
and derived the UV and infrared luminosities of
individual galaxies by post-processing with a radiative transfer code.
\citet{Bekki:2015aa} treated dust as a separated particle
type from gas, dark matter and stellar components in a smoothed particle
hydrodynamics (SPH) simulation, and computed dust formation in stellar ejecta,
dust growth by accretion, and dust destruction in SN shocks.
\citet{McKinnon:2016ab} implemented dust formation and destruction
in a moving-mesh code {\small AREPO},
in which dust is treated as an attribute of gaseous fluid element.
Their cosmological zoom-in simulations revealed that
dust growth by accretion is important,
and that appropriate stellar and AGN feedback models
are necessary to reproduce the observed dust-to-metal ratios at
$z=0$.
Furthermore, \citet{McKinnon:2017aa} performed cosmological simulations with
the same framework but additionally considering sputtering
in the hot gas environment.
Their dust mass function and cosmic comoving dust density are consistent with
observations in the local Universe,
but their simulation has a tendency of underestimating
the number of dust-rich galaxies at high redshift.
\citet{Zhukovska:2016aa} examined the effect of dust growth and destruction
in multiphase ISM by post-processing an SPH simulation.
They studied temperature-dependent sticking coefficient
in the accretion of gas-phase metals onto dust.
For the above simulations, however, an important caveat is that
the grain size distribution is not taken into account, since,
as mentioned above, it has a large
influence on extinction curves and dust evolution.

\citet{Aoyama:2017aa} implemented the dust enrichment model
in an SPH simulation of an isolated galaxy.
Their dust model includes dust production in stellar ejecta,
dust destruction in SN shocks, dust growth by accretion,
grain growth by coagulation, and grain disruption by shattering.
The last two processes are caused by grain--grain collisions
and are important in determining the grain size distribution.
The grain size distribution is represented by
the abundances of `large' and `small' grains separated
at $a\sim 0.03~\mu$m,
following the two-size approximation by \citet{Hirashita:2015aa}.
This simplification serves to calculate the grain size information
within a reasonable computational time.
\citet{Hou:2017aa} further separated the dust species into
silicate and carbonaceous dust, and found that
the combination of grain size distribution
and grain species in the simulation allows us
to calculate the spatially-resolved extinction curves.
They succeeded in reproducing the Milky Way extinction curve
in solar-metallicity environments,
and predicted the density and metallicity dependence of extinction curve,
which could produce a dispersion of extinction curves
in various lines of sight.

In this paper, we further investigate statistical properties
of dust evolution in a cosmological volume.
\citet[][hereafter Paper I]{Aoyama:2018aa}
extended the framework mentioned above
to a cosmological simulation.
They studied overall dust properties in a cosmological volume and
in the intergalactic medium (IGM) without entering the details of
individual galaxies.
This paper aims to study various dust scaling relations in galaxies
and their evolution.
We examine two basic dust property indicators,
dust-to-gas ratio and dust-to-stellar mass ratio,
against galaxy properties such as metallicity,
stellar mass, gas fraction and specific SFR (sSFR).
Grain size distribution and extinction curves
are also studied in this work.

This paper is organised as follows.
In Section~\ref{section:method}, we describe
the cosmological simulation with dust enrichment.
We present the dust scaling relations, the redshift evolution of
those relations and the extinction curves in Section~\ref{section:result}.
In Section~\ref{section:discussion},
we discuss possible improvements of our models
and make predictions on galaxies at $z >$ 5.
Finally, we provide the conclusions in Section~\ref{section:conclusion}.
We adopt the cosmological parameters according to
\citet{Planck-Collaboration:2016aa}:
baryon density parameter $\Omega_{\rm b}$ = 0.049,
total matter density parameter $\Omega_{\rm m}$ = 0.32,
cosmological constant parameter $\Omega_{\Lambda}$ = 0.68,
Hubble constant $H_0$ = 67 km s$^{-1}$ Mpc$^{-1}$,
power spectrum index $n_{\rm s}$ = 0.9645,
and density fluctuation normalisation $\sigma_{\rm 8}$ = 0.831.
We also use non-dimensional Hubble constant $h \equiv
H_0 / (100\,{\rm km\,s^{-1}\,Mpc^{-1}}$).
For the consistency with our pervious papers, we adopt
$Z_{\sun}=0.02$ for the solar metallicity.


\section{Model}
\label{section:method}

In this section, we describe the simulation,
the dust evolution model, and the galaxy identification method.
These basically follow Paper I, but there are some differences:
First, we implement a simple
AGN feedback model in the simulation to improve the prediction on
massive galaxies. Second, we consider the
silicon and carbon as the key materials of grain growth by accretion
to refine the treatment of accretion.
Third, we consider different SN destruction efficiency
between large and small grains since small grains are more easily
destroyed. The latter two points do
not cause significant differences from Paper I.
We also describe the calculation method of extinction curves,
which are new in this paper.

\subsection{Cosmological simulation}
\label{model:cosmo_simulation}

The modified version of \textsc{gadget}-3 $N$-body/SPH code
\citep[last described in][]{Springel:2005aa}
was used for this study.
The initial number of particles are $N = 2 \times~ 512^3$ (gas and dark matter),
and the comoving simulation box size is 50 $h^{-1}$\,Mpc.
We refer to the gas SPH particles as the gas particles in this paper.
The CELib chemical evolution library \citep{Saitoh:2017aa} and
the Grackle\footnote{https://grackle.readthedocs.org/}
chemistry and cooling library \citep{Smith:2017aa}
were implemented
(Shimizu et al., submitted).
The chemical enrichment includes not only
instantaneous metal injection from Type II SNe
but also delayed metal production of Type Ia SNe and AGB stars.

The Chabrier initial mass function (IMF) \citep{Chabrier:2003ab}
from 0.1 to 100 $\Msun$ is adopted.
Star formation is allowed only for the gas particles
with $n_{\rm gas} \ge 0.1~{\rm cm}^{-3}$ and $T_{\rm gas} < 10^4$ K,
where $n_{\rm gas}$ and $T_{\rm gas}$ are the number density
and gas temperature, respectively.
The star formation efficiency is modified to a slightly lower value
($\varepsilon_\mathrm{SF}=0.01$)
from Paper I ($\varepsilon_\mathrm{SF}=0.05$)
based on the comparison against the observations of the
local Universe (see Fig.\ \ref{Fig:MF_star}) after the inclusion of
the AGN feedback described below.
Stellar feedback is treated consistently with metal production (Shimizu et al. submitted).
The basic setup of simulation is summarised in Table~\ref{table:simulation}.

\begin{table}
\caption[Simulation setup.]
{Simulation setup.
$N$, $\varepsilon_{\rm grav}$, $m_{\rm dm}$, and $m_{\rm gas}^{\rm init}$
are the number of particles, the gravitational softening length,
the mass of dark matter particle and the initial mass of gas particle, respectively.}
\begin{center}
\begin{tabular}{ccccc}
\hline
Boxsize & $N$ & $\varepsilon_{\rm grav}$ & $m_{\rm dm}$ & $m_{\rm gas}^{\rm init}$ \\
{[$h^{-1}$\,Mpc]} &  & [$h^{-1}$\,kpc] & [$h^{-1}\,\Msun$] & [$h^{-1}\,\Msun$] \\
\hline
$50$ & $2 \times 512^3$ & $3$ & $6.89 \times 10^7$ & $1.28\times 10^7$ \\
\hline
\end{tabular}
\label{table:simulation}
\end{center}
\end{table}

We have found that Paper I tends to overproduce the metallicity of
massive galaxies. Although this overproduction does not affect the
statistical analysis in the cosmological volume in Paper I, it may affect
our results focusing on individual galaxies. Therefore, we newly attempt to
include suppression of metal enrichment (or star formation) in massive
halos by the so-called AGN feedback.
AGN feedback is known to be important in galaxy evolution especially
for massive galaxies \citep[e.g.][]{Booth:2009aa,Vogelsberger:2014aa}.
Many studies pointed out that AGN feedback is significant
in reproducing high-mass end of galaxy mass function
\citep[e.g.][]{Di-Matteo:2005aa,Springel:2005ab,Croton:2006aa,Booth:2009aa,Harrison:2017aa}.
Here we introduce a phenomenological AGN feedback model based on
\citet{Okamoto:2014aa}, who assumed that radiative cooling is inefficient
in massive galaxies where one-dimensional dark matter velocity dispersion is greater than
the following threshold, $\sigma_\mathrm{th}$, as a function of $z$:
\begin{equation}
\sigma_{\rm th}(z) = \sigma_0 (1 + z)^{\alpha}.
\end{equation}
Here $\sigma_0$ is the normalisation parameter and
$\alpha$ controls the redshift dependence.
We set $\sigma_0 = 100~{\rm km/s}$ and $\alpha = 0.75$ \citep{Okamoto:2014aa}.
In this study, we adopt sudden suppression of gas cooling
above $\sigma_\mathrm{th}$ rather than the
smooth suppression model used in the original paper,
because of a higher affinity to
the Grackle cooling routine.
Fortunately, as we show in this paper, there is no unwanted bump
around the mass corresponding to $\sigma_{\rm th}$
in the galaxy stellar mass function which is seen in the original paper.

\subsection{Identification of galaxies}

We first identify dark matter halos by the Friends-of-Friends grouping algorithm
with a linking length of 0.2 times the mean dark matter particle separation
\citep{Davis:1985aa}.
Following  \citet{Nagamine:2004aa} and \citet{Choi:2009aa},
we identify galaxies based on baryonic components using the
\textsc{subfind} algorithm \citep{Springel:2001aa}.
This method first computes the smoothed density field for baryonic
particles to locate the centre of individual galaxies
with isolated density peaks.
The galaxy is constructed by adding the star and gas particles
one by one in the order of declining density.
If all the 512 nearest neighbour particles have lower densities,
this particle is considered to be a new galaxy seed.
Otherwise, the particle is attached to the galaxy which
the nearest denser neighbour particle belongs to.
If the two nearest denser neighbour particles are in different
galaxies and one of the galaxies have particle number less than 32,
two galaxies are merged into one.
On the other hand, if the two nearest denser neighbour particles
belong to different galaxy and both of them have
particle number greater than 32, this particle is assigned to the
larger one.
Each galaxy has to include at least 32 particles; otherwise
it is not recognized as a galaxy.
We only analyse galaxies with stellar mass $M_\ast > 10^8\,\Msun$.

\subsection{Galaxy properties in the simulation}\label{subsec:simulation_result}

\begin{figure}
	\begin{center}
	\includegraphics[width=0.475\textwidth]{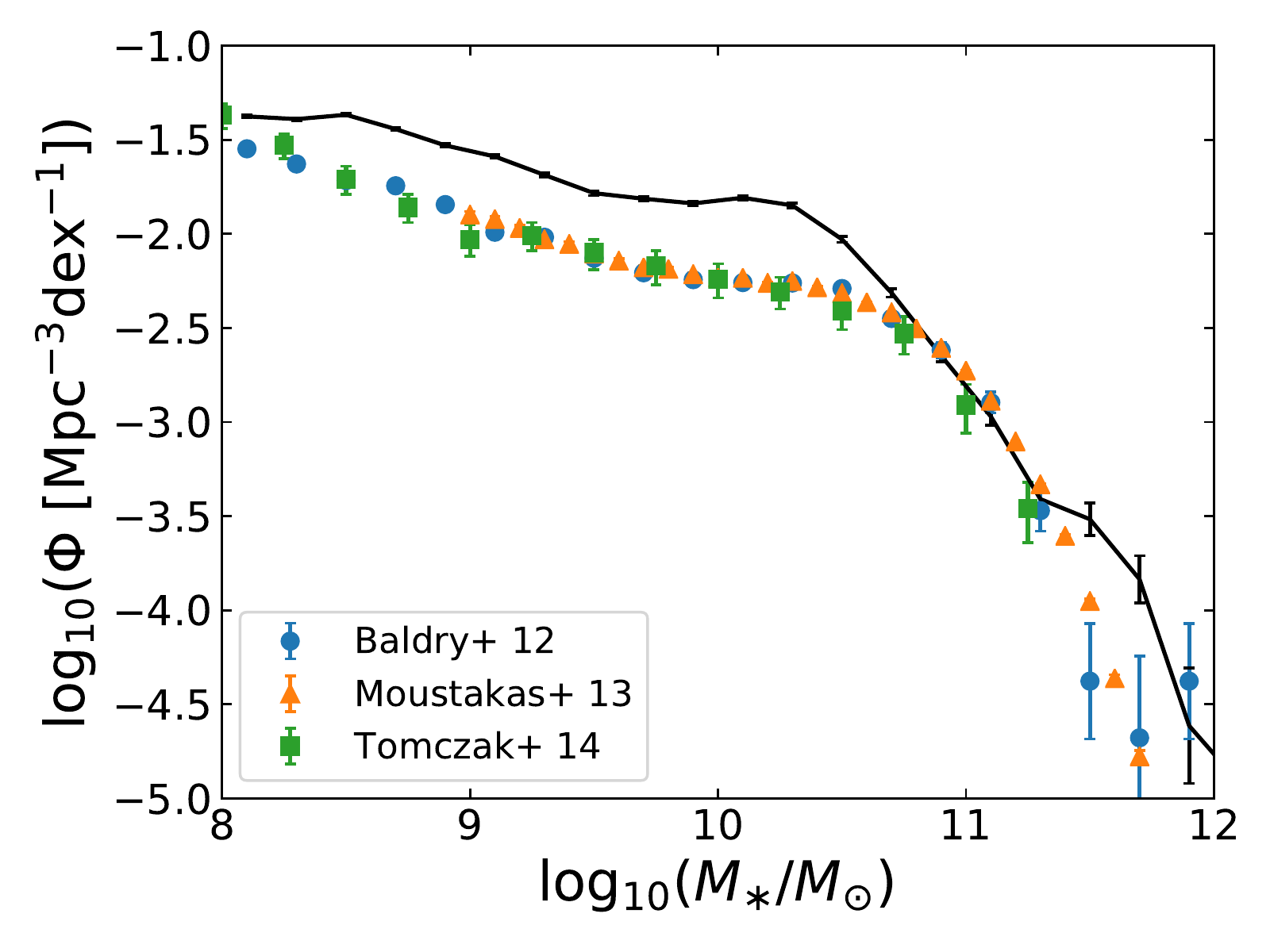}
	\caption{Galaxy stellar mass function at $z = 0$
	(solid line with Poisson error bars).
	We over-plot the observational data from
	\citet[][circles]{Baldry:2012aa}, \citet[][triangles]{Moustakas:2013aa},
	and \citet[][squares]{Tomczak:2014aa}.
	}
	\label{Fig:MF_star}
	\end{center}
\end{figure}

\begin{figure}
	\begin{center}
	\includegraphics[width=0.475\textwidth]{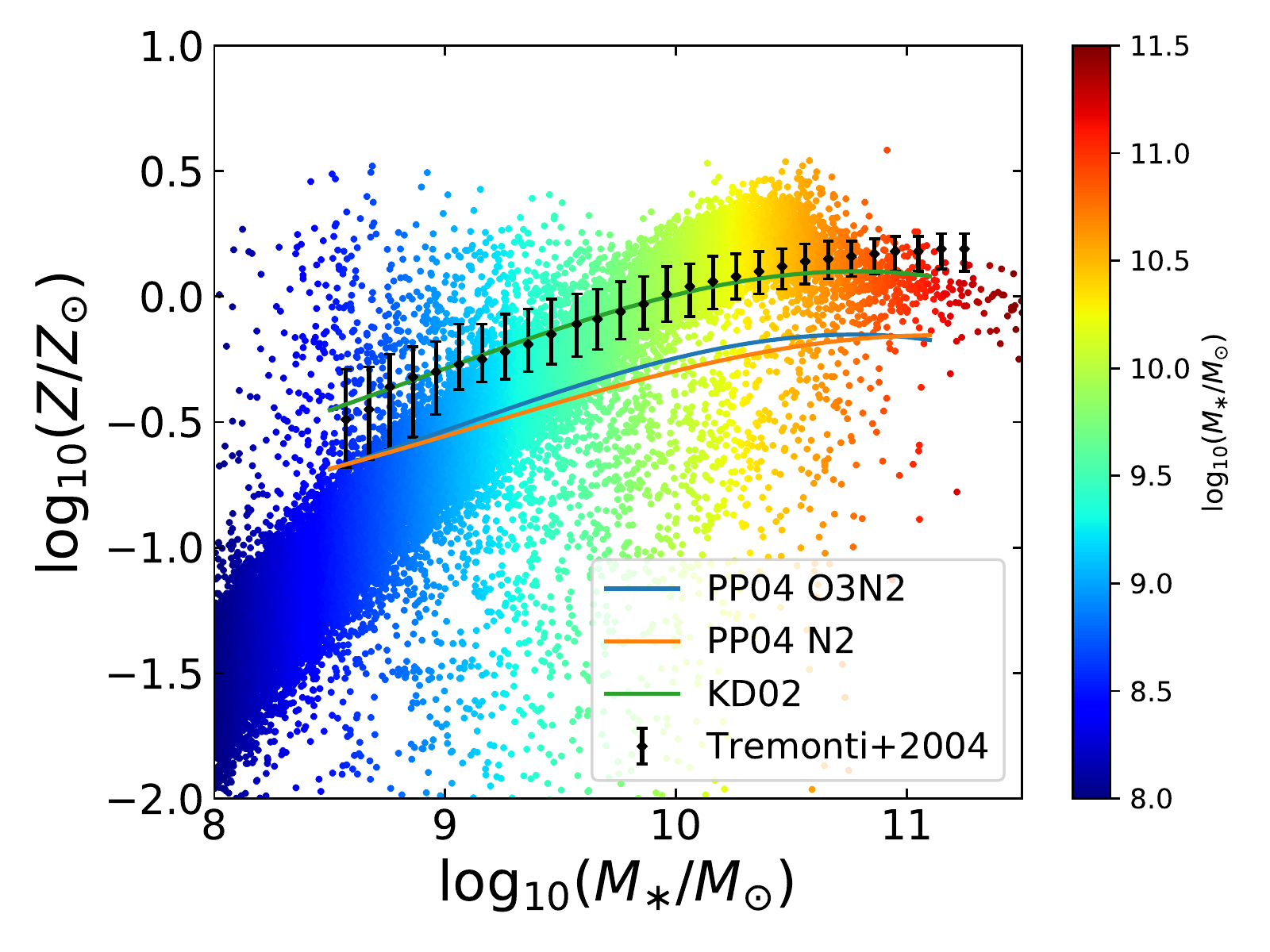}
	\caption{Relation between gaseous metallicity and stellar mass at $z = 0$.
	Each point represents a galaxy with the colour indicating the stellar mass
	as shown in the colour bar.
	The black points with errors are observational data
	in the local Universe from \citet{Tremonti:2004aa}.
	The blue, orange and green lines show the best-fit relations for star-forming galaxies
	in SDSS using various metallicity calibrators \citep{Kewley:2008aa}.
	PP04 O3N2, PP04 N2 and KD02 in the legend correspond to metallicity indicators proposed by
	\citet{Pettini:2004aa} and \citet{Kewley:2002aa}.
	}
	\label{Fig:Ms_Z}
	\end{center}
\end{figure}

Figure~\ref{Fig:MF_star} shows the galaxy stellar mass function
in our simulation at $z=0$.
Compared with Paper I, the newly added simple AGN feedback
reduces the galaxy number density at the high-mass end; yet,
we slightly overproduce the number of galaxies at the massive
end compared with the observations
\citep{Baldry:2012aa,Moustakas:2013aa,Tomczak:2014aa}.
The mass function is also overproduced below the knee.
However, these discrepancies between the model and the observations
are within a factor of $\sim 2$.
On the other hand, as we show in Fig. \ref{Fig:Ms_Z},
the relation between
stellar mass and gas-phase metallicity
(hereafter, we simply refer to this relation as stellar mass--metallicity
relation) is reasonably reproduced with our model.
Compared with Paper I, we lower the star formation efficiency to
$\varepsilon_{\rm SF} = 0.01$ 
to better reproduce the stellar mass--metallicity relation,
which reduces the strength of stellar feedback and causes the
over-prediction of galaxy number density at the low-mass end.
If we strengthen the stellar feedback in low-mass galaxies, we could
reproduce the stellar mass function, but we would significantly underproduce
the metallicity.
Since metallicity has a direct
influence on dust evolution, the agreement with the
stellar mass--metallicity relation is more important than that
to the stellar mass function; thus, we adopt the current model.
Moreover, since our AGN feedback model is simple, it is not possible to obtain a
perfect fit to all data. The factor 2 uncertainties do not significantly affect the
discussed trends in the scaling relations below.

Since dust evolution is strongly related to metal enrichment,
we examine whether the chemical enrichment is
described reasonably in terms of stellar mass growth
in our simulation.
Figure~\ref{Fig:Ms_Z} shows the $M_{\ast}$--$Z$ relation
($Z$ is the gas metallicity) at $z=0$.
The relation is consistent with the observation
\citep{Tremonti:2004aa} at
$M_\ast\gtrsim 10^9$ M$_{\sun}$.
We confirmed that, if we do not include AGN feedback,
we overproduce the metallicity compared with observational data
at the high-$M_\ast$ end.
Thus, our simple AGN feedback model succeeded in suppressing the
chemical enrichment in massive galaxies.
At lower masses of $M_\ast \lesssim 10^{8.5}$ M$_{\sun}$, there is not a strong
observational constraint, but there seems to be a tendency
that the simulation underestimates the metallicity.
The underproduction of $Z$ in low-mass galaxies
is probably due to insufficient resolution of the simulation; that is,
we are unable to resolve dense gas, which leads to
an underestimate of the cooling rate especially in low-mass galaxies at low redshifts
and accordingly, to an underestimate of star formation activity.
Thus, the discussions in this paper put a focus on galaxies with $M_\ast\gtrsim 10^9\,\Msun$.


\subsection{Dust enrichment model}

Basically, we use the same dust enrichment model as
in \citet{Aoyama:2017aa} and \citet{Hou:2017aa}. The model is
based on the two-size dust enrichment model developed by \citet{Hirashita:2015aa}.
In this model, the grain size distribution is represented by only two sizes,
`large' and `small' grains, divided at $a\sim 0.03~\micron$ ($a$ is the grain radius).
For dust evolution processes, we consider dust production in
stellar ejecta, grain destruction by SN shock,
grain disruption by shattering in the diffuse ISM,
and grain growth by coagulation and accretion in the dense ISM.
In Paper I, we considered delayed stellar dust production
by Type Ia SNe and AGB stars in addition to the instantaneous contribution from
Type II SNe.
We also include sputtering in the hot circum-galactic medium (CGM) and IGM.
However, this newly included sputtering is unimportant for this paper,
since we focus on the dust in the ISM.

We use the large grain dust-to-gas ratio
$\mathcal{D}_{{\rm L}} \equiv m_{\rm d,L}/m_{\rm gas}$
and small grain dust-to-gas ratio
$\mathcal{D}_{{\rm S}} \equiv m_{\rm d,S}/m_{\rm gas}$
to formulate the dust evolution,
where $m_{\rm d,L/S}$ and $m_{\rm gas}$ are the large/small dust mass
and the gas mass of each gas particle, respectively.
The total dust-to-gas ratio is defined as
$\mathcal{D} \equiv \mathcal{D}_{{\rm L}} + \mathcal{D}_{{\rm S}}$.
The time evolution of the large- and small-grain dust-to-gas ratios in each particle
from time $t$ to the next time step $t + \Delta t$ is formulated with following equations:
\begin{align}
\mathcal{D}_{{\rm L}}(t+\Delta t)
&= \mathcal{D}_{{\rm L}}(t) - \Delta \mathcal{D}^{\rm SN}_{{\rm L}}(t)
+ f_{\rm in}\dfrac{\Delta m_{{\rm metal}}}{m_{\rm gas}} ( 1 - \delta )~ \notag \\
&+ \left(- \dfrac{\mathcal{D}_{{\rm L}}(t)}{\tau_{\rm sh}}
+ \dfrac{\mathcal{D}_{{\rm S}}(t)}{\tau_{\rm co}}
- \dfrac{\mathcal{D}_{{\rm L}}(t)}{\tau_{\rm sp}(a_{\rm L})}\right)\Delta t
- \mathcal{D}_{{\rm L}}(t)\dfrac{\Delta m_{\rm gas}^{\rm return}}{m_{\rm gas}}\, ,
\label{eq:large} \\
\mathcal{D}_{{\rm S}}(t+\Delta t)
&=\mathcal{D}_{{\rm S}}(t) - \Delta \mathcal{D}^{\rm SN}_{{\rm S}}(t)
- \mathcal{D}_{{\rm S}}(t)\dfrac{\Delta m_{\rm gas}^{\rm return}}{m_{\rm gas}}\notag \\
&+ \left(\dfrac{\mathcal{D}_{{\rm L}}(t)}{\tau_{\rm sh}}
- \dfrac{\mathcal{D}_{{\rm S}}(t)}{\tau_{\rm co}}
+ \dfrac{\mathcal{D}_{{\rm S}}(t)}{\tau_{\rm acc}}
- \dfrac{\mathcal{D}_{{\rm S}}(t)}{\tau_{\rm sp}(a_{\rm S})}\right)\Delta t\, ,
\label{eq:small}
\end{align}
where $\Delta \mathcal{D}^{\rm SN}_{\rm L/S}(t)$
is the decrease by the SN destruction of pre-existing dust,
$f_{\rm in}$ is the dust condensation efficiency in stellar ejecta,
$\Delta m_{{\rm metal}}$ is the ejected metal mass,
$\Delta m_{\rm gas}^{\rm return}$ is the gas ejection rate,
$\delta$ is the fraction of newly formed dust that is destroyed by SNe.
The time-scale parameters,
$\tau_{\rm sh}$, $\tau_{\rm co}$, $\tau_{\rm acc}$ and $\tau_{\rm sp}$,
are for shattering, coagulation, accretion and sputtering, respectively
(see below).
The representative grain radii for large and small grains are
$a_{\rm L} = 0.1$\,$\mu$m and $a_{\rm S} = 0.005$\,$\mu$m, respectively.

In this two-size approximation,
stellar dust production is the source of large grains,
shattering converts large grains into small grains,
coagulation transforms small grains into large grains, and
accretion increases only the small grain abundance.
Sputtering in SNe and hot gas decreases
both small and large grain abundances.
Stellar dust production is treated consistently with
metal production assuming that the fraction $f_\mathrm{in}$ of
the newly formed metals condense into dust.
The value of $f_\mathrm{in}$ is in the range of $\sim0.01 - 0.5$
which varies among theoretical models adopted
\citep{Inoue:2011aa,Kuo:2013aa,Hou:2016aa}, and it was treated
as a parameter in the previous theoretical calculations \citep{Hirashita:2015aa} and
simulations \citep{Aoyama:2017aa}. Following Paper I, we adopt $f_\mathrm{in} = 0.1$
in this work.

We derive $\mathcal{D}$ for each galaxy
by summing up all the dust mass and gas mass in gas particles
with $T_\mathrm{gas}\leq 5 \times 10^4$\,K
contained in the galaxy and taking their ratio.
The temperature cut here is to eliminate the contamination from the CGM
since we are interested in the dust in galaxies, not in halos.
Since our estimate of $\mathcal{D}$ is based on a mass-weighted average,
$\mathcal{D}$ reflects the value in the dense part (i.e.\ the ISM) of each galaxy.
We tested various threshold gas temperatures between $10^4$ and $10^6$\,K
and confirmed that results are not sensitive to the selected value.
We have clarified this in Section 2.4.
The small-to-large grain abundance ratio, $\stol$, is also derived 
by taking the ratio between the total small-grain mass and large-grain mass of the galaxy.

\subsubsection{SN destruction}\label{subsubsec:SN}

SNe destroy dust grains in their sweeping radius.
Because each stellar particle contains more than $\sim 10^6$ stars,
we consider a sub-grid model to represent multiple SN explosions,
and separate destructions between pre-existing dust and newly formed dust
to avoid double-counting SN destruction.

The destruction of pre-existing dust can be written as
\begin{equation}
{\Delta \mathcal{D}^{{\rm SN}}_{{\rm L\slash S}} (t)}=
\left[ 1-( 1 - \eta )^{N_\mathrm{SN}}\right] {\mathcal{D}^{{\rm SN}}_{{\rm L\slash S}}(t)},
\end{equation}
where $\eta \equiv\min [\epsilon_\mathrm{SN}(m_\mathrm{sw}/m_\mathrm{gas}),\,\epsilon_\mathrm{SN}]$
($m_\mathrm{sw}$ is the gas mass swept by a single SN estimated by
\citealt{Aoyama:2017aa},
and $\epsilon_\mathrm{SN}$ is the efficiency of dust destruction in a single SN blast),
and $N_\mathrm{SN}$ is the number of SN explosions in the gas particle.
Since small grains are more easily destroyed in SN shocks than large grains
\citep[e.g.][]{Nozawa:2006aa}, we adopt
$\epsilon_\mathrm{SN}=0.1$ for large grains and
$\epsilon_\mathrm{SN}=1$ for small grains (while Paper I applied
$\epsilon_\mathrm{SN}=0.1$ for both grain populations).
The fraction of the newly formed dust that survives after a passage of $N_\mathrm{SN}$
SN shocks is
\begin{equation}
(1 - \delta) = \frac{1}{N_\mathrm{SN}} \frac{1 - ( 1 - \eta)^{N_\mathrm{SN}}}{\eta}.
\end{equation}
The derivation was described in Appendix A of \citet{Aoyama:2017aa}.

\subsubsection{Shattering}
Shattering is assumed to occur in low-density regions (with number density
$n_{\rm gas} < 0.1$\,cm$^{-3}$),
where the characteristic grain velocity is expected to be high enough for shattering.
The shattering time-scale is estimated as
\begin{equation}
\tau_\mathrm{sh} = 5.41 \times 10^{8}\ \mathrm{yr}
\left(\frac{n_{\rm gas}}{1~\mathrm{cm}^{-3}}\right)^{-1}
\left( \dfrac{\mathcal{D}_{\rm L}}{0.01} \right)^{-1},
\label{tau_sh}
\end{equation}
For gas particles with $n_\mathrm{gas} \ge 0.1 ~\mathrm{cm}^{-3}$,
we turn off shattering.

\subsubsection{Coagulation and accretion}

Coagulation occurs in dense regions where grain velocities are low.
In the two-size approximation, coagulation is regarded as a process in which
two small grains stick to form a larger grain. Thus, the coagulation time-scale
is determined by the collision time between small grains.
Accretion is a process in which small grains gain their mass
by accreting gas-phase metals,
so the accretion time-scale has a metallicity dependence.
For accretion, we newly consider key species (silicon and carbon;
see equation \ref{eq:tau_acc} below)
while Paper I considered all the metals equally.
However, this change does not cause any significant differences in
the resulting dust abundance.

Dense clouds are important for accretion and coagulation, but
they cannot be fully resolved in our simulation.
Therefore, we adopt the following sub-grid treatment:
10 per cent of the gas mass is in the state of dense cloud
(with density and temperature $10^3$\,cm$^{-3}$ and $50$\,K,
respectively)
if the gas density is greater than $n_{\rm gas} = 0.1$\,cm$^{-3}$.

We only consider coagulation and accretion in dense and cold gas particles satisfying
$n_\mathrm{gas} \ge 0.1 ~\mathrm{cm}^{-3}$
and $T_\mathrm{gas} < 10^4$ K.
The time-scale of coagulation is estimated as
\begin{equation}
\tau_\mathrm{co} = 2.70 \times 10^{5}\ \mathrm{yr}\
\left( \dfrac{\mathcal{D}_{\rm S}}{0.01} \right)^{-1}\slash f_{\rm dense},
\label{tau_co}
\end{equation}
and the time-scale of accretion is formulated as
\begin{equation}
\tau_\mathrm{acc} = 1.05 \times 10^{6}\ \mathrm{yr}
\left(\dfrac{Z_{\rm C+ 6 Si}}{Z_\odot} \right)^{-1}
\left(1-\dfrac{\mathcal{D}}{Z_{\rm C+ 6 Si}}\right)^{-1}\slash f_{\rm dense},
\label{eq:tau_acc}
\end{equation}
where $Z_{\rm C+ 6 Si}$ is the abundance of dust-composing metals
(see below)
and $f_{\rm dense}$ is the fraction of dense cloud ($f_\mathrm{dense}=0.1$
following Paper I).
We assume that Si is the key element for silicate, and that the mass
fraction of Si in silicate is 1/6 \citep{Draine:1984aa}.
We suppose that carbonaceous dust is purely composed of C.
Therefore, we calculate the carbon abundance plus 6 times the silicon
abundance to obtain $Z_\mathrm{C+6Si}$ and use it as the abundance of
dust-composing material.
We adopt $f_{\rm dense} = 0.1$.

\subsubsection{Sputtering}

We also include dust destruction by sputtering in the hot gas not
associated with SNe (note that we have already treated dust destruction
in SN shocks in Section \ref{subsubsec:SN}).
Such hot gas mainly exists in the CGM and IGM.
To avoid double-counting the SN destruction,
we extract only the hot gas component not associated with SNe
by imposing the gas density limit
$n_\mathrm{gas}<0.01$\,cm$^{-3}$, which is typical for the CGM and IGM.
We consider sputtering only in gas particles with $n_\mathrm{gas}<0.01$\,cm$^{-3}$
and $T_\mathrm{gas}>10^6$ K.
We adopt the following destruction time-scale based on \citet{Tsai:1995aa}:
\begin{equation}
\tau_\mathrm{sp}(a) = 2.15 \times 10^{5}\, \mathrm{yr}
\left(\dfrac{a}{1\, \mu{\rm m}}\right)
\left(\dfrac{n_{\rm gas}}{1\, {\rm cm}^{-3}}\right)^{-1}.
\label{eq:tau_sp}
\end{equation}


\subsection{Extinction curves}
\label{subsec:extinc_curve}

The grain size distribution could be observationally tested by
extinction curves.
The calculation of extinction curve in our paper is based on
the method described by \citet{Hou:2016aa}.

Because we need to assume dust compositions, we apply
the solar elemental pattern to determine the relative fraction of
silicate and carbonaceous dust. We denote the abundances
of silicate and carbonaceous dust as $\mathcal{D}_\mathrm{Si}$
and $\mathcal{D}_\mathrm{C}$, respectively. The abundance of
each species is estimated as
$\mathcal{D_{\rm X}} = \mathcal{D} \times F_\mathrm{X}(Z_{\rm X\odot}/Z_\odot)$,
where subscript X represents the dust species (X = C or Si) and
$Z_{\rm C\odot} = 2.47 \times 10^{-3}$ and $Z_{\rm Si\odot} = 8.17 \times 10^{-4}$
are the solar carbon and silicon abundances, respectively.
The factor $F_\mathrm{X}$ is introduced to account for the elements
other than Si and C. We adopt $F_\mathrm{Si}=6$ and $F_\mathrm{C}=1$.

To calculate the extinction curve, the grain size distribution
is required; however, the grain size distribution is represented at
only two sizes in our simulation.
Thus, we need to assume a specific functional form for the grain
size distribution of each population.
Following \citet{Hirashita:2015aa}, we adopt a modified lognormal function
for the grain size distribution:
\begin{equation}
  n_{i,\mathrm{X}}(a)=\frac { C_{i,\mathrm{X}} }{ a^{ 4 } }
  \exp\left\{ - \frac { \left[ \ln(a/a_{ 0,i }) \right] ^{ 2 } }{ 2\sigma^2 }  \right\},
\end{equation}
where subscript $i$ indicates small ($i = \mathrm{S}$) or large
($i = \mathrm{L}$) grain component,
$C_{i,\mathrm{X}}$ is the normalisation constant,
$a_{0,i}$ and $\sigma$ are the central grain radius
and the standard deviation of the lognormal part, respectively.
We adopt $a_\mathrm{0,S} = 0.005\,\micron$,
$a_\mathrm{0,L} = 0.1\,\micron$ and $\sigma = 0.75$
since these values reproduce the Milky Way extinction curve
when $\stol$ is the same as the \citet[][MRN]{Mathis:1977aa}
size distribution \citep{Hirashita:2015aa}.
The normalisation $C_{i,\mathrm{X}}$ is determined by
\begin{equation}
  {\mu}m_\mathrm{H}\mathcal{D}_{i,\mathrm{X}}=
  \int_{0}^{\infty}\frac{4}{3}{\pi}a^3\rho_\mathrm{X}n_{i,\mathrm{X}}(a)\,\mathrm{d}a,
\end{equation}
where $\mu = 1.4$ is the gas mass per hydrogen nucleus,
$\rho_{\rm X}$ is material density
($\rho_{\rm Si} = 3.5$ and $\rho_{\rm C} = 2.24$\,g\,cm$^{-3}$
for silicate and carbonaceous dust, respectively)
and $m_\mathrm{H}$ is the mass of hydrogen atom.

The extinction $A_{\lambda,\mathrm{X}}$ (in units of magnitude) normalised
to the column density of hydrogen nuclei ($N_\mathrm{H}$) is written as
\begin{equation}
  \frac{A_{\lambda,\mathrm{X}}}{N_\mathrm{H}}=
  2.5\,(\log_{10} \mathrm{e})\, \sum_{i}\int_{0}^{\infty}
  n_{i,\mathrm{X}}(a){\pi}a^2Q_\mathrm{ext}(a,\lambda,\mathrm{X})\,\mathrm{d}a,
\end{equation}
where 
$Q_\mathrm{ext}(a,\lambda,\mathrm{X})$ is the extinction coefficient
(extinction cross-section normalised to the geometric cross-section) as a
function of grain radius, wavelength and dust species.
$Q_\mathrm{ext}(a,\lambda,\mathrm{X})$ is calculated by the Mie theory
\citep{Bohren:1983aa} based on the same optical constants for silicate and
graphite as in \citet{Weingartner:2001aa}. The carbonaceous dust is represented
by graphite in this paper, but we note that we may have to consider other
carbonaceous materials to explain bumpless extinction curves
\citep{Nozawa:2015aa,Hou:2016aa}.
Since we are interested in the extinction curve, we always normalize
the extinction to $A_V$ (extinction in the $V$ band). Thus,
$N_\mathrm{H}$ cancels out in the final plots.
Using the above equations,
we calculate the mean extinction curve for each galaxy based on
the resulting value of $\stol$.


\section{Results}
\label{section:result}

We investigate the statistical properties of dust content in this section.
We also examine the relation between a dust-related quantity
of individual
galaxies and principal galaxy characteristics;
namely, stellar mass, gas mass and SFR, so that
we can investigate scaling relations
regarding dust.
First, we study the statistical properties of galaxies in the local Universe.
In later subsections, we show the redshift evolution.

\subsection{Dust mass function}
\label{result:DMF}

\begin{figure}
	\begin{center}
	\includegraphics[width=0.475\textwidth]{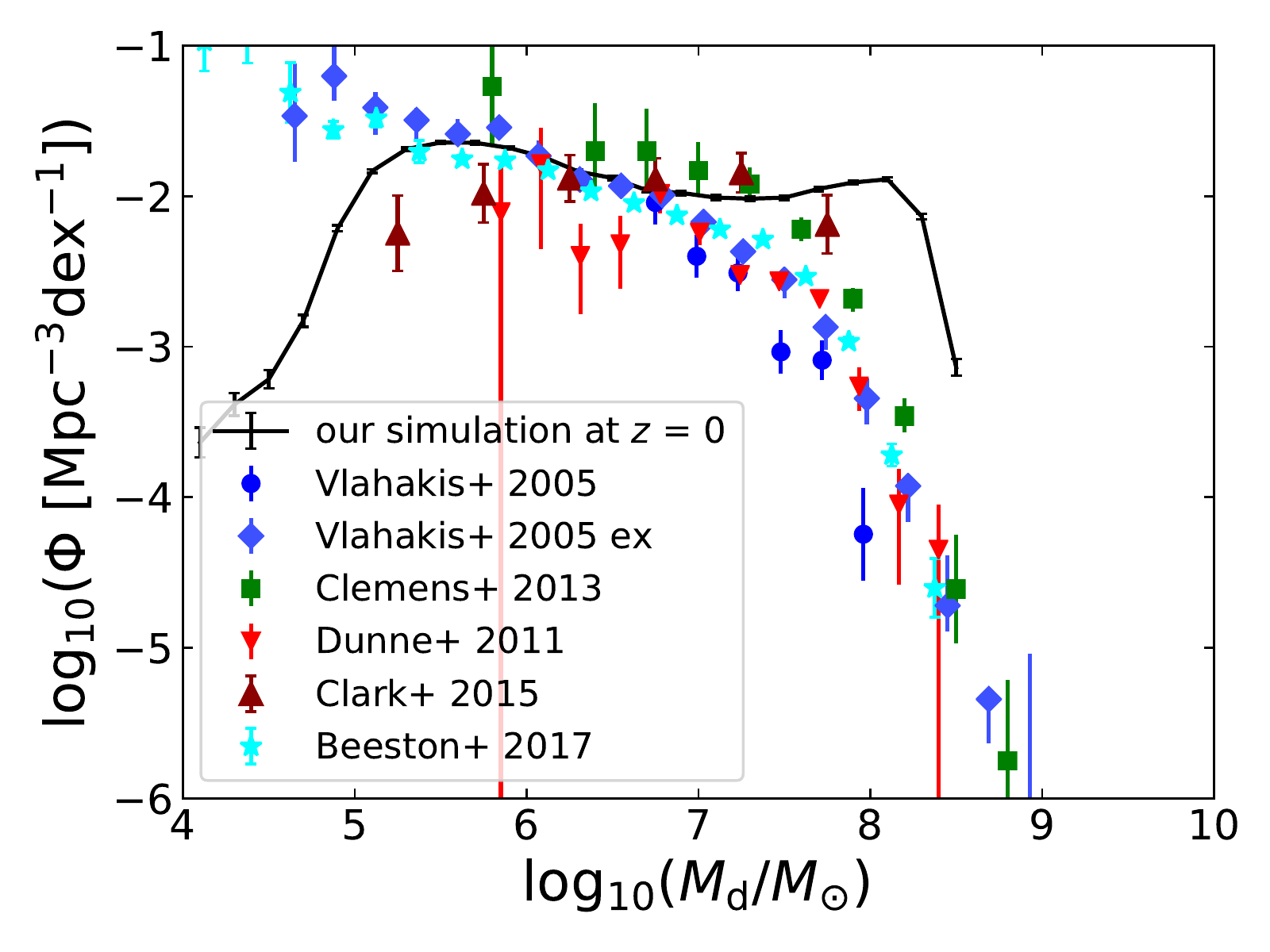}
	\caption{Dust mass function at $z = 0$.  Our simulation result
	is shown by the solid line with Poissonian errors.
	Observational data points are taken from \citet[][including the PSCz-extrapolated
	data denoted by `Vlahakis+ 2005 ex']{Vlahakis:2005aa},
	\citet{Dunne:2011aa}, \citet{Clemens:2013aa}, \citet{Clark:2015aa},
	and \citet{Beeston:2017aa} as shown in the legend.
	}
	\label{Fig:DustMF}
	\end{center}
\end{figure}

The statistics of galaxy dust mass can be represented by the dust mass function,
that is, the distribution function of dust mass in galaxies.
We show the dust mass function at $z=0$ in Fig.~\ref{Fig:DustMF}.
For comparison, we compile the observational data of dust mass function in the local Universe
\citep{Vlahakis:2005aa,Dunne:2011aa,Clemens:2013aa,Clark:2015aa,Beeston:2017aa}.
\citet{Vlahakis:2005aa} derived the local dust mass function for
optically selected SCUBA 850\,$\micron$ sources supplemented by
\textit{IRAS} Point Source Catalogue Redshift Survey (PSCz) catalog.
They also provided the dust mass function for submillimetre non-detected sources
by extrapolating the SEDs of the PSCz galaxies to longer wavelengths.
We denote this extrapolated estimate in the legend with a suffix `ex' in
Fig.~\ref{Fig:DustMF}.
\citet{Dunne:2011aa} derived the dust mass function from
\textit{Herschel} 250\,$\micron$ sources with the Sloan Digital Sky Survey (SDSS)
counterparts for $z< 0.5$.
\citet{Clemens:2013aa} combined the \textit{Planck} data with
those taken by the infrared space telescopes and fitted the SEDs
to derive the dust mass function within a distance of 100\,Mpc.
\citet{Clark:2015aa} obtained the dust mass function from
a volume-limited sample (between distances 15 and 46\,Mpc)
in \textit{Herschel} Astrophysical Terahertz Large Area Survey
(\textit{H}-ATLAS) with SDSS counterparts.
\citet{Beeston:2017aa} derived a dust mass function
by SED fitting in the overlapping fields between the
Galaxy and Mass Assembly (GAMA) and \textit{H}-ATLAS.
For all the above estimates,
we re-evaluate the dust mass by a common mass absorption coefficient
at 850\,$\micron$ as $\kappa_{850} = 0.77$\,cm$^2$\,g$^{-1}$
\citep{Vlahakis:2005aa,Clark:2015aa}
with assumed wavelength dependence of
$\kappa_{\lambda} \propto \lambda^{-2}$.
As shown in \citet{Hirashita:2014aa},
the estimated dust mass is uncertain
by $\sim 0.5$\,dex because of the uncertainty in the mass absorption coefficient.

Our simulation reproduces the dust mass function at dust masses
$10^5 \lesssim M_\mathrm{d} \lesssim 10^{7.5}\,\Msun$.
At the high-mass end ($M_{\rm d}>10^{7.5}\,\Msun$), our simulation
overproduces the dust mass function.
As shown later in Section \ref{Result:D2G}, we reproduce the relation
between dust-to-gas ratio and metallicity, which means that we do not
overproduce the dust abundance relative to the metallicity.
In addition, as discussed in Section \ref{subsec:simulation_result},
the metallicity is not significantly overproduced for a given stellar mass.
Compared with the observational data in \citet{Saintonge:2017aa},
our galaxies have a few times higher gas-to-stellar mass ratio
($M_{\rm gas}/M_\ast$).
The gas mass in our simulated galaxies also has an excess
compared with the ALFALFA survey results from \citet{Maddox:2015aa}
at the high stellar mass end ($M_\ast > 10^{10} M_\odot$).
This leads to a high dust mass at the massive end.

Thus, our simple AGN feedback model does not blow out the gas
efficiently in massive galaxies.
However, not only the AGN feedback but also star formation and stellar feedback
processes are tightly related to the gas fraction in galaxies.
Since our main goal is not to study the feedback processes, we leave this issue for the future work.
Nevertheless, we still confirm that the galaxies with $M_\mathrm{d}>10^{8.5}\,\Msun$,
which existed in Paper I, do not appear in our simulation any more because of
the newly implemented AGN feedback model.

\citet{McKinnon:2017aa} performed cosmological simulations
using a moving-mesh code {\small AREPO} with a dust enrichment model.
Their dust mass function is higher at
$M_{\rm d} \lesssim 10^{6}$\,$\Msun$ than ours,
because they adopted higher dust condensation efficiencies (0.5 to 1)
depending on the dust species and source (AGB or SN)
\citep{McKinnon:2016ab}.
Note that the dust condensation efficiency is $f_\mathrm{in}=0.1$ in our simulation.
Their method also has a difference in that they did not enhance the gas density by
adopting a sub-grid model; thus,
dust growth by accretion is much weaker in their simulation than in ours.
This less efficient dust growth could be a reason why their
dust mass function is at the massive end is successfully suppressed
compared to ours.
A semi-analytic model with dust formation and destruction by
\citet{Popping:2017aa} also showed an over-prediction of the dust mass
function at the high-mass end ($M_{\rm d} \gtrsim 10^8$\,$\Msun$),
although they also included the AGN feedback in their model.


\subsection{Dust-to-gas ratio}
\label{Result:D2G}

The dust-to-gas ratio, $\mathcal{D}$, is a fundamental quantity for dust evolution.
We first investigate the
relation between $\mathcal{D}$ and other galaxy properties
such as metallicity ($Z$), stellar mass ($M_{\ast}$), sSFR
($\equiv$ SFR/$M_\ast$)
and gas fraction ($f_{\rm gas}$)
at $z = 0$.

\begin{figure*} 
	\begin{center}
	\includegraphics[width=0.475\textwidth]{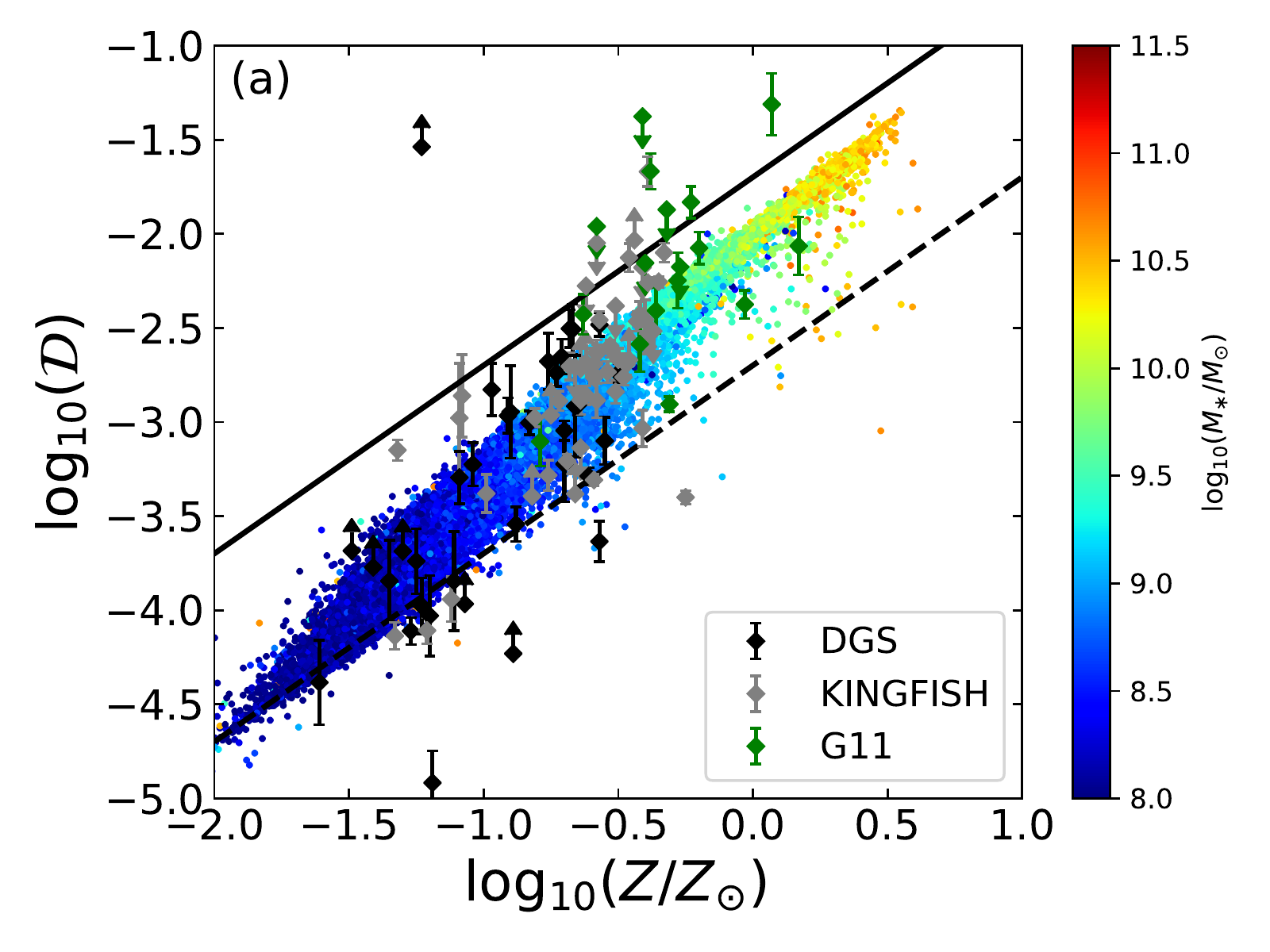}
	\includegraphics[width=0.475\textwidth]{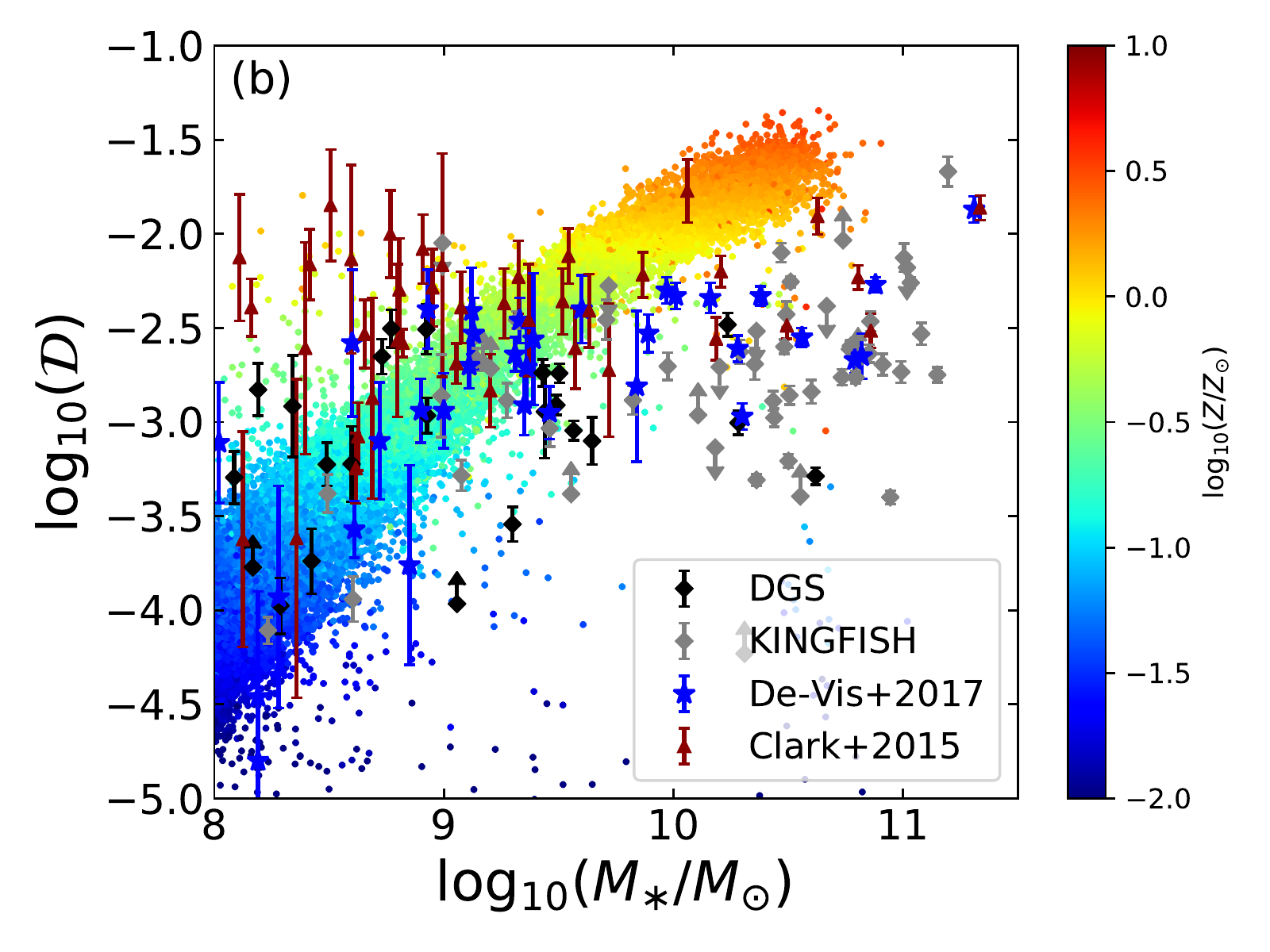}
	\includegraphics[width=0.475\textwidth]{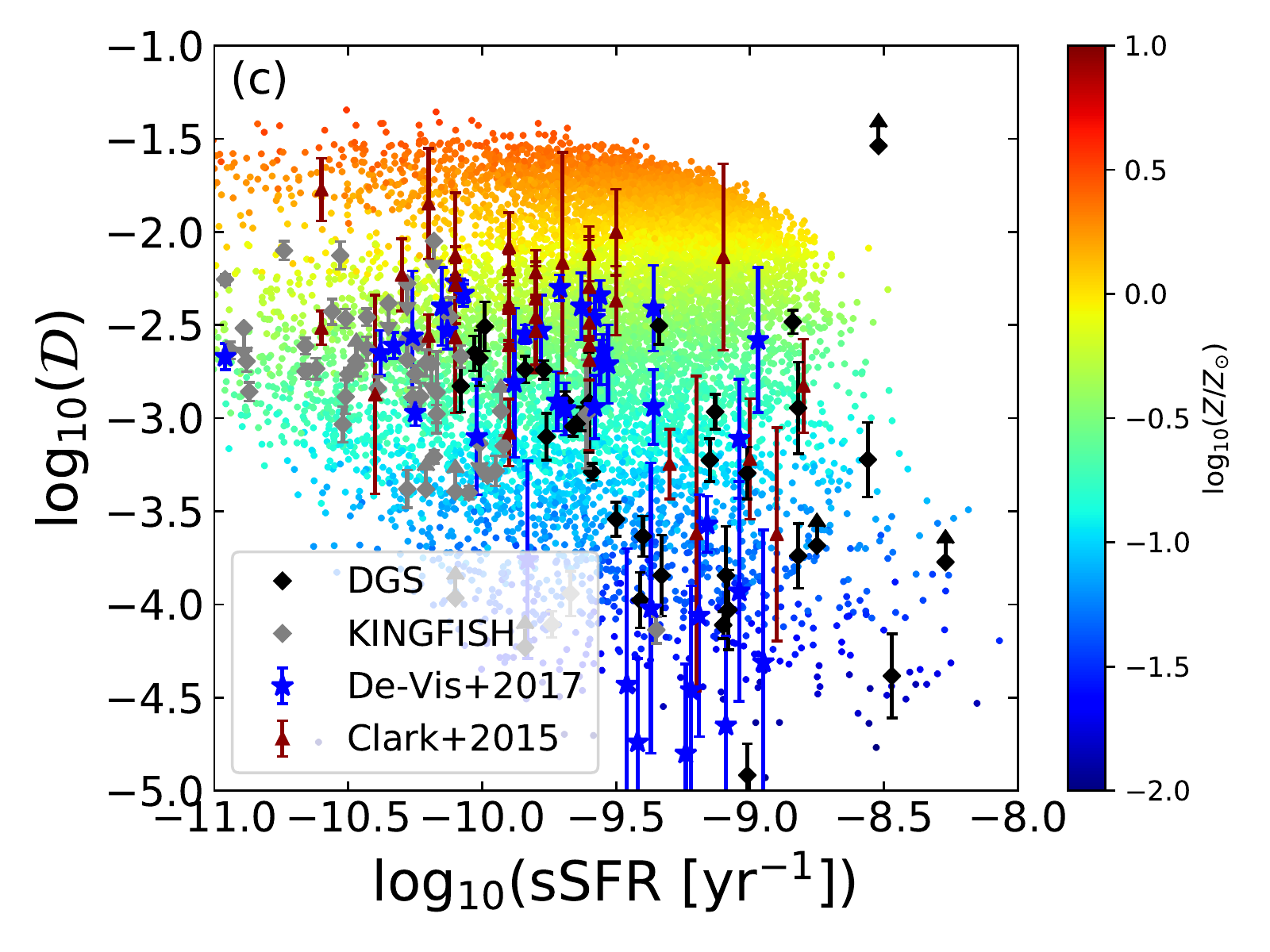}
	\includegraphics[width=0.475\textwidth]{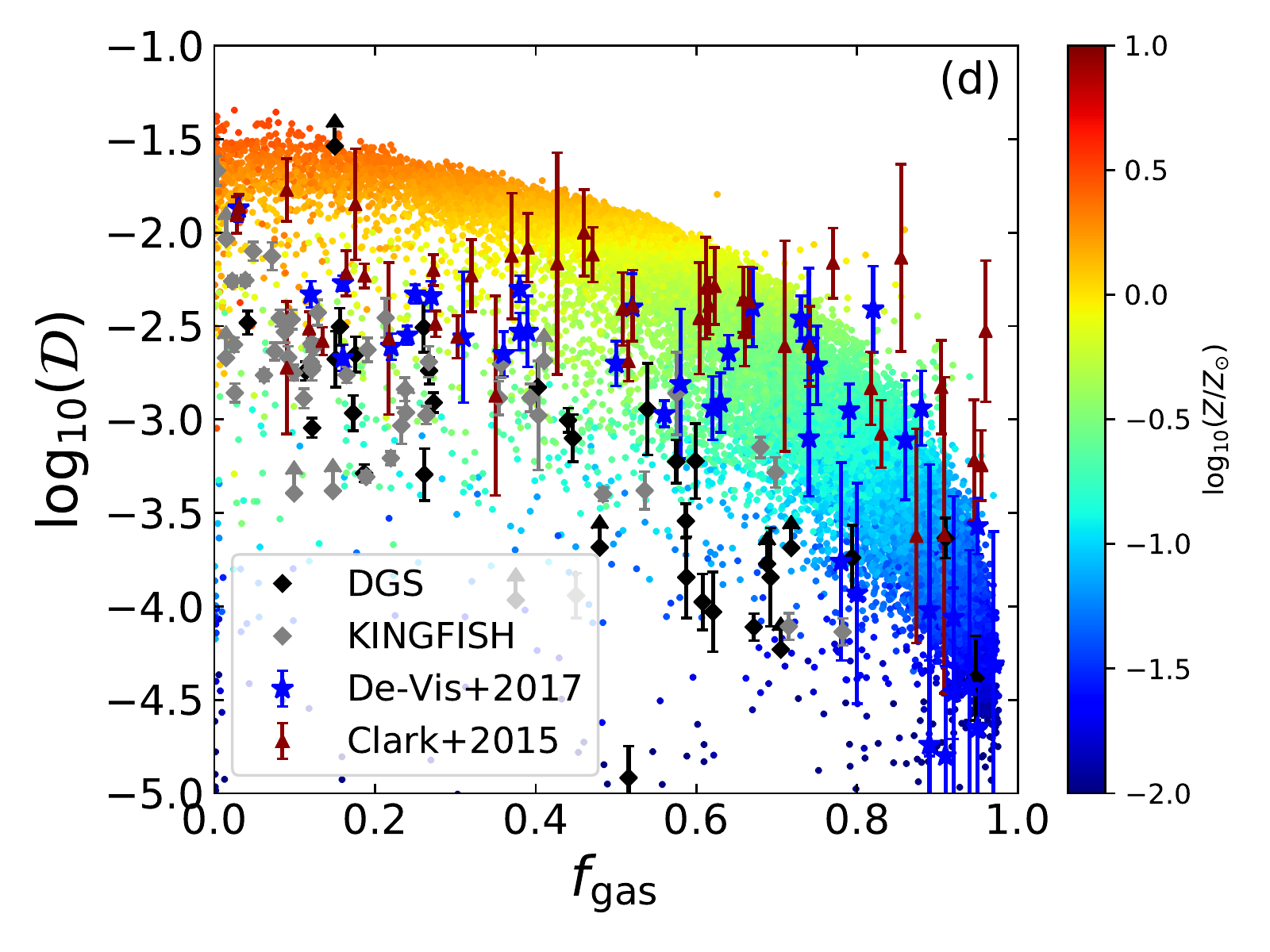}
	\caption{Scaling relations of dust-to-gas ratio at $z = 0$ against
	(a) metallicity, (b) stellar mass, (c) sSFR,
	and (d) gas fraction.
	Each point represents a galaxy with its colour indicating the stellar mass
	(in panel {a}) or metallicity (panels {b, c, and d}) as shown in the colour bar.
	The solid and dashed lines in panel ({a}) show the saturation
	limit ($\mathcal{D_{\rm tot}} = Z$) and the linear relation of the stellar yield
	($\mathcal{D_{\rm tot}} = f_{\rm in}Z$), respectively. The
	observational points are taken from \citet{Remy-Ruyer:2014aa} (G11, DGS and KINGFISH),
	\citet{Clark:2015aa}, and \citet{De-Vis:2017aa} as shown in the legend.
	}
	\label{Fig:D2G}
	\end{center}
\end{figure*}

In Fig.~\ref{Fig:D2G}a, we show the $\mathcal{D}$--$Z$ relation.
Since dust evolution is mainly driven by metal enrichment,
$\mathcal{D}$ has a positive relation to $Z$.
At $Z \lesssim 0.05~Z_\odot$, the $\mathcal{D}$--$Z$ relation is determined by
the stellar dust production. As a consequence, the
relation follows a linear relation $\mathcal{D}\simeq f_{\rm in}Z$
at low metallicity.
There is a steep nonlinear increase of $\mathcal{D}$ between
$\sim 0.05~Z_\odot$ and $\sim 0.5~Z_\odot$ because of dust growth by accretion.
At $Z \gtrsim 0.5~ Z_\odot$,
$\mathcal{D}$ approaches $Z$ as dust growth is saturated.
We also show nearby star-forming galaxy data taken from
\citet[][]{Remy-Ruyer:2014aa} (DGS, KINGFISH and G11 samples;
see their paper for the detailed description of the samples).
They collected galaxies with a large
variety in metallicity. They derived the dust mass from the FIR SED and
the gas mass from the sum of H\,\textsc{i} and H$_2$ (converted from CO).
The H\,\textsc{i} mass is corrected to match the extent of the dust emission.
Our result shows good agreement with the observational data,
including the nonlinear increase by dust growth.
This nonlinear relation was also shown by other cosmological
evolution models \citep{de-Bennassuti:2014aa,Popping:2017aa}.
There are some observational data points with $\mathcal{D}$
higher than $Z$.  However, $\mathcal{D}$ must not exceed $Z$ by definition.
Those data points might be affected by large observational uncertainties in
estimating the dust abundance,
and we do not aim to explain them using our simulation.

In Fig.~\ref{Fig:D2G}b, we show the $\mathcal{D}$--$M_{\ast}$ relation.
The relation is flatter at $M_{\ast} \gtrsim 10^{9}$\,$\Msun$ than
at $M_{\ast} \lesssim 10^{9}$\,$\Msun$.
This transition of slope basically follows the relation between
metallicity and stellar mass (Fig.~\ref{Fig:Ms_Z});
thus, the $\mathcal{D}$--$M_{\ast}$ relation is governed by chemical enrichment.
The stellar mass is truncated at $M_{\ast} \sim 10^{10.7}\,\Msun$ in this figure:
because galaxies beyond this mass are strongly affected by AGN feedback,
they lack the cold gas phase containing
dust (recall that we only counted the dust mass in gas particles satisfying
the temperature criterion, $T_\mathrm{gas} < 5 \times 10^4$ K).
Therefore, our simple AGN feedback model succeeds in suppressing
the dust abundance at the massive end, and potentially explains
the poor dust content in massive elliptical galaxies.
For comparison,
we overplot the nearby galaxy sample in \citet{Remy-Ruyer:2014aa}.
In addition, we also adopt
other nearby-galaxy data taken from \citet{Clark:2015aa} and \citet{De-Vis:2017aa};
both papers used the sample
in the \textit{H}-ATLAS, and
the former and latter studies are based on dust-selected and H\,\textsc{i}-selected
samples, respectively.  Note that the
gas mass in these samples only include the H\,\textsc{i} mass.
There are 22 overlapping sources between these two papers;
however, the dust masses obtained for the same object are not necessarily
the same,
because \citet{De-Vis:2017aa} re-analysed the \textit{Herschel} photometry,
and adopted a different SED fitting technique from that used by \citet{Clark:2015aa}.
Thus, we plot full results from both data without removing overlapping galaxies.
Since there are no stellar mass data for the G11 sample,
only DGS and KINGFISH are shown here for \citet{Remy-Ruyer:2014aa}'s data.
The observational data do not show strong correlation between $\mathcal{D}$
and $M_{\ast}$.
The dust-to-gas ratios in our simulation
are higher than the observed ones at
$M_{\ast} \gtrsim 10^{10}$\,$\Msun$, while
our results at $M_{\ast} \lesssim 10^{9}$\,$\Msun$ are well
within the observed large scatter in the dust-to-gas ratio.
The excess at the massive end could be partly
due to an underestimate of AGN feedback in our model.
However, we should note that we do not overproduce the metallicity in
this stellar mass range significantly
(as we show in Fig.~\ref{Fig:Ms_Z},
there could be a factor 2 over-production of metallicity
around $M_\ast\sim 10^{10}$ M$_{\sun}$), and that the relation between
dust-to-gas ratio and metallicity is consistent with observations even at
the high-metallicity end as shown above.
Even if we suppress the dust-to-gas ratio by a factor 2, there still
seems to be a overproducing tendency in the dust-to-gas ratio
at $M>10^{9.5}$ $\Msun$. We should also point out the uncertainty
in the observational estimate of dust-to-gas ratio. Indeed, some
high-$\mathcal{D}$ objects in Fig.\,\ref{Fig:D2G}a do not appear in
the observational sample in Fig.\,\ref{Fig:D2G}b, which implies that
there is some bias in the data.

Fig.~\ref{Fig:D2G}c presents the $\mathcal{D}$--sSFR relation.
In our simulation,
galaxies with high sSFR tend to have low $\mathcal{D}$,
although there is a large scatter.
The number of low-metallicity ($Z\lesssim 0.1$ Z$_{\sun}$) galaxies
is less in this panel than in the others because
a large fraction of these low-metallicity galaxies, which are mostly low-mass
galaxies, form stars intermittently and have no SFR in the snapshot at $z=0$
(all the other plots regarding sSFR also have
less low-metallicity points).
The observational data taken from the same papers
as above
\citep{Remy-Ruyer:2014aa,Clark:2015aa,De-Vis:2017aa}
show the trend of lower sSFR for higher $\mathcal{D}$,
which is correctly reproduced by our simulation.
We notice that there are some high-metallicity galaxies that have high
sSFR since an excessive
amount of gas still remains in these systems as mentioned above in
Section \ref{result:DMF}.

In Fig.~\ref{Fig:D2G}d, we plot the $\mathcal{D}$--$f_{\rm gas}$ relation, where
the gas fraction $f_{\rm gas}$ is defined as $M_{\rm gas}/(M_{\rm gas} + M_{\ast})$.
We find that galaxies with low $f_{\rm gas}$ tend to have high $\mathcal{D}$.
This trend is produced because chemical enrichment proceeds as more gas is
converted to stars.
We plot the same observational samples as used in Fig.\ \ref{Fig:D2G}b.
The decreasing trend of $\mathcal{D}$ for increasing $f_\mathrm{gas}$ is
consistent with the observational data. The scatter in the simulation data is
also comparable to that in the observational data.

The above results show that our modeling of dust-to-gas ratio correctly
reproduces the relation with other quantities, except that the dust-to-gas
ratio may be overproduced at $M_\ast\gtrsim 10^{10}$ M$_{\sun}$. On the
other hand, we should note uncertainties and possible bias
against dust-rich objects in the
observational samples as pointed out above.


\subsection{Dust-to-stellar mass ratio}
\label{Result:D2S}

Dust-to-stellar mass ratio, $M_{\rm d}/M_{\ast}$, is also an important quantity
to understand dust enrichment in terms of stellar mass growth.
Since stellar emission is usually easier to
observe than gas emission, it is sometimes useful to
qualitatively investigate $M_{\rm d}/M_{\ast}$ rather than $\dtog$.
In the following, we compare $M_{\rm d}/M_{\ast}$ of simulated galaxies
with the same quantities as in the previous subsections
($Z$, $M_{\ast}$, sSFR and $f_{\rm gas}$).

\begin{figure*}
	\begin{center}
	\includegraphics[width=0.475\textwidth]{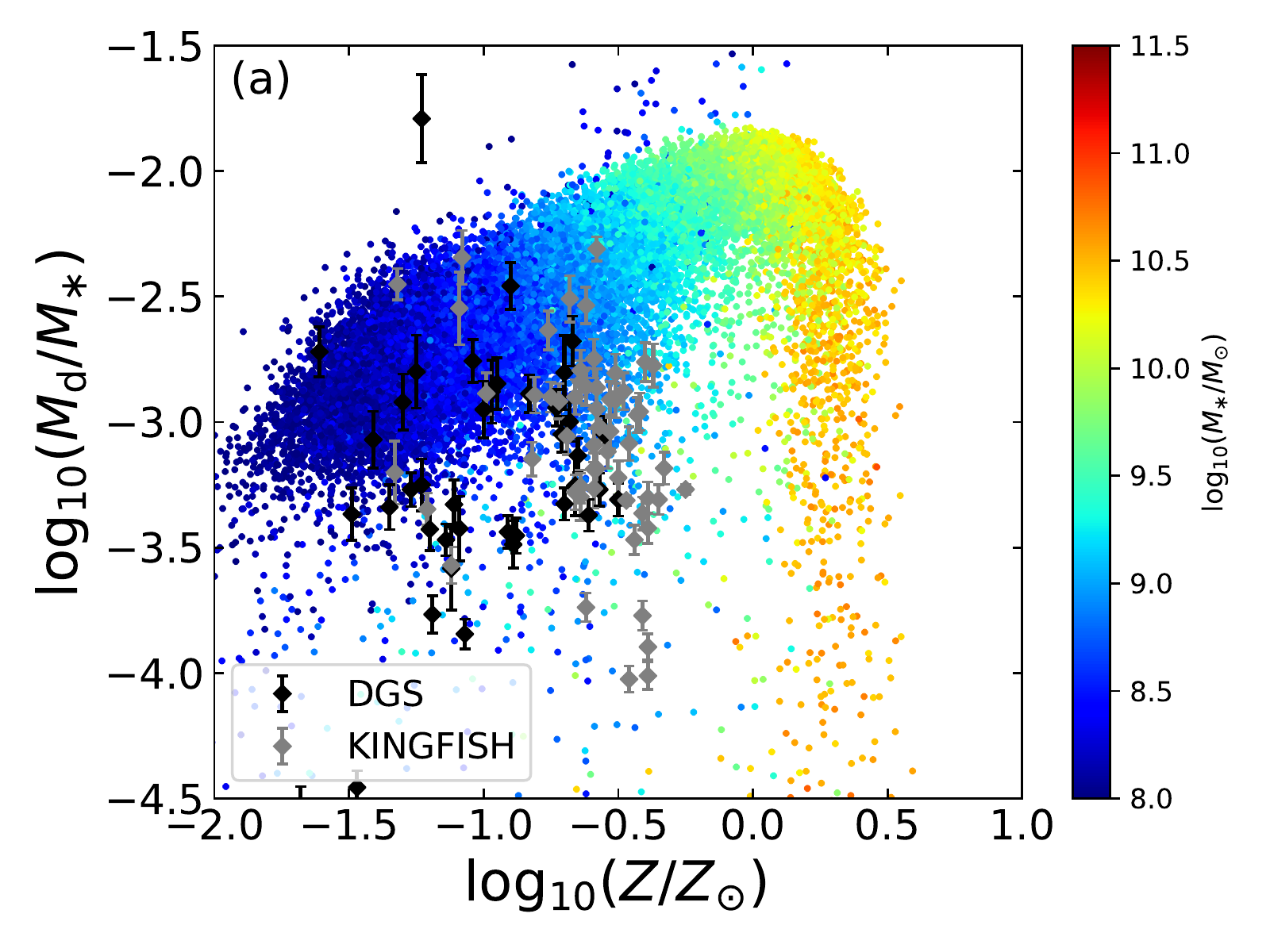}
	\includegraphics[width=0.475\textwidth]{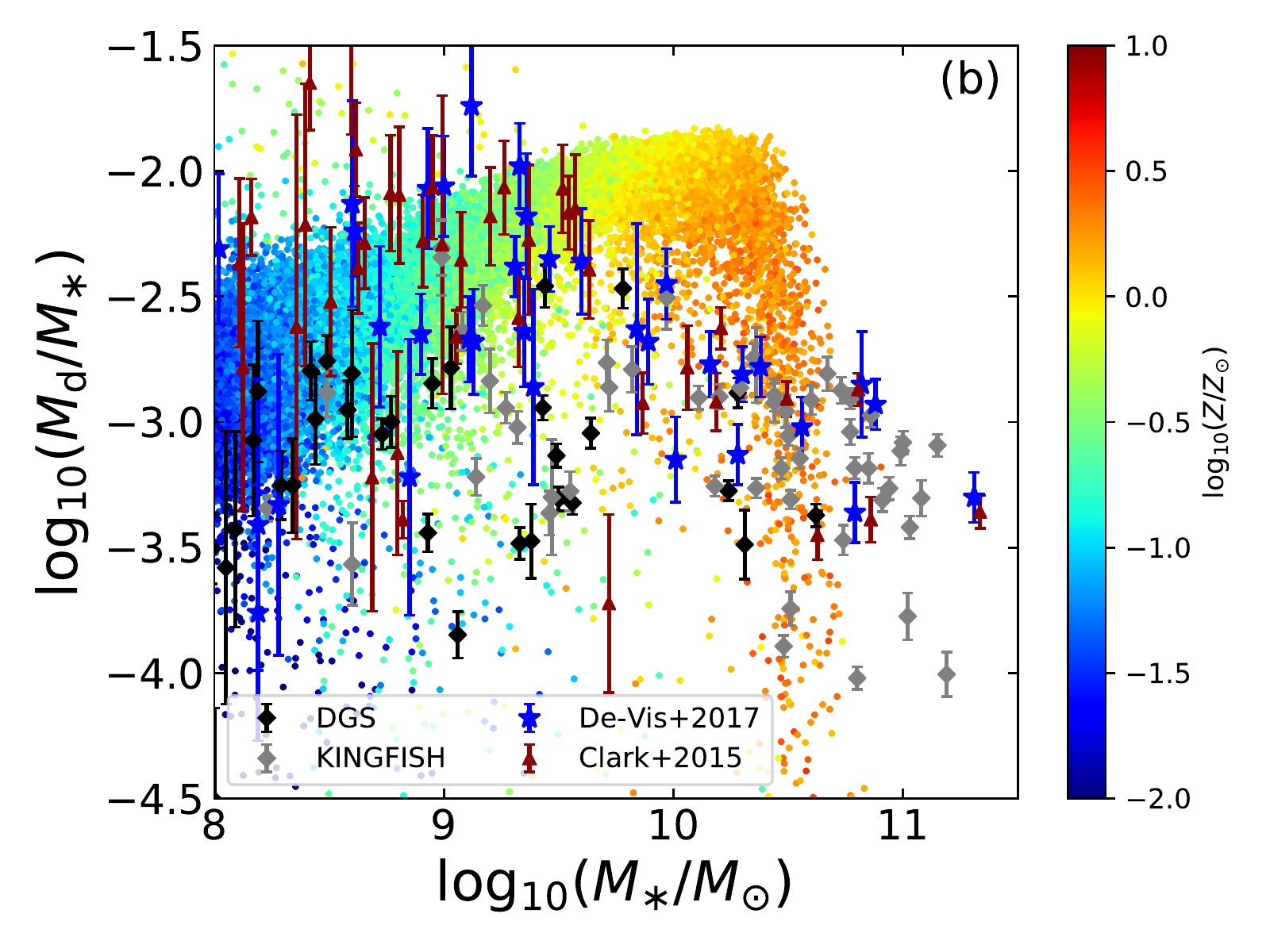}
	\includegraphics[width=0.475\textwidth]{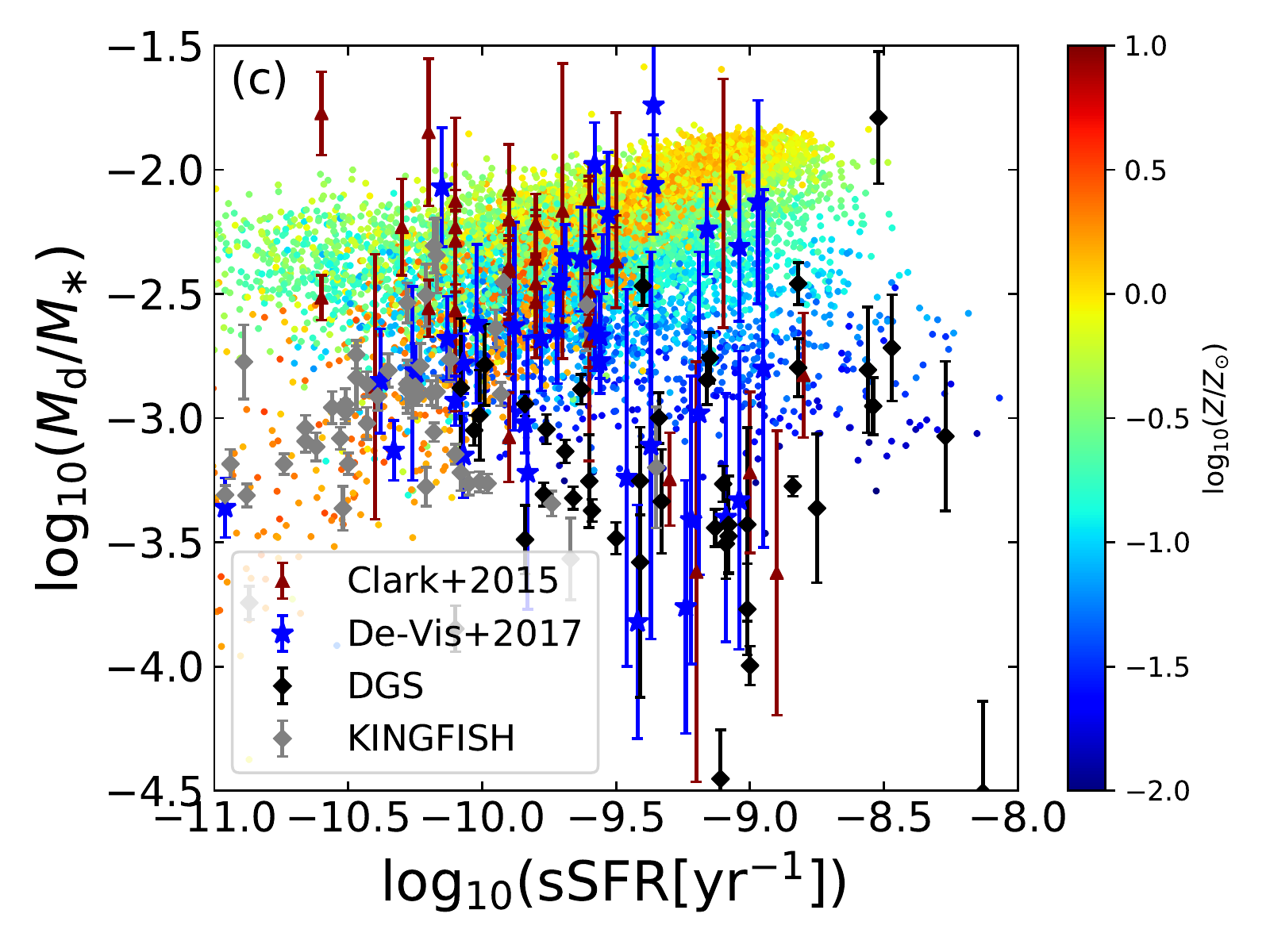}
	\includegraphics[width=0.475\textwidth]{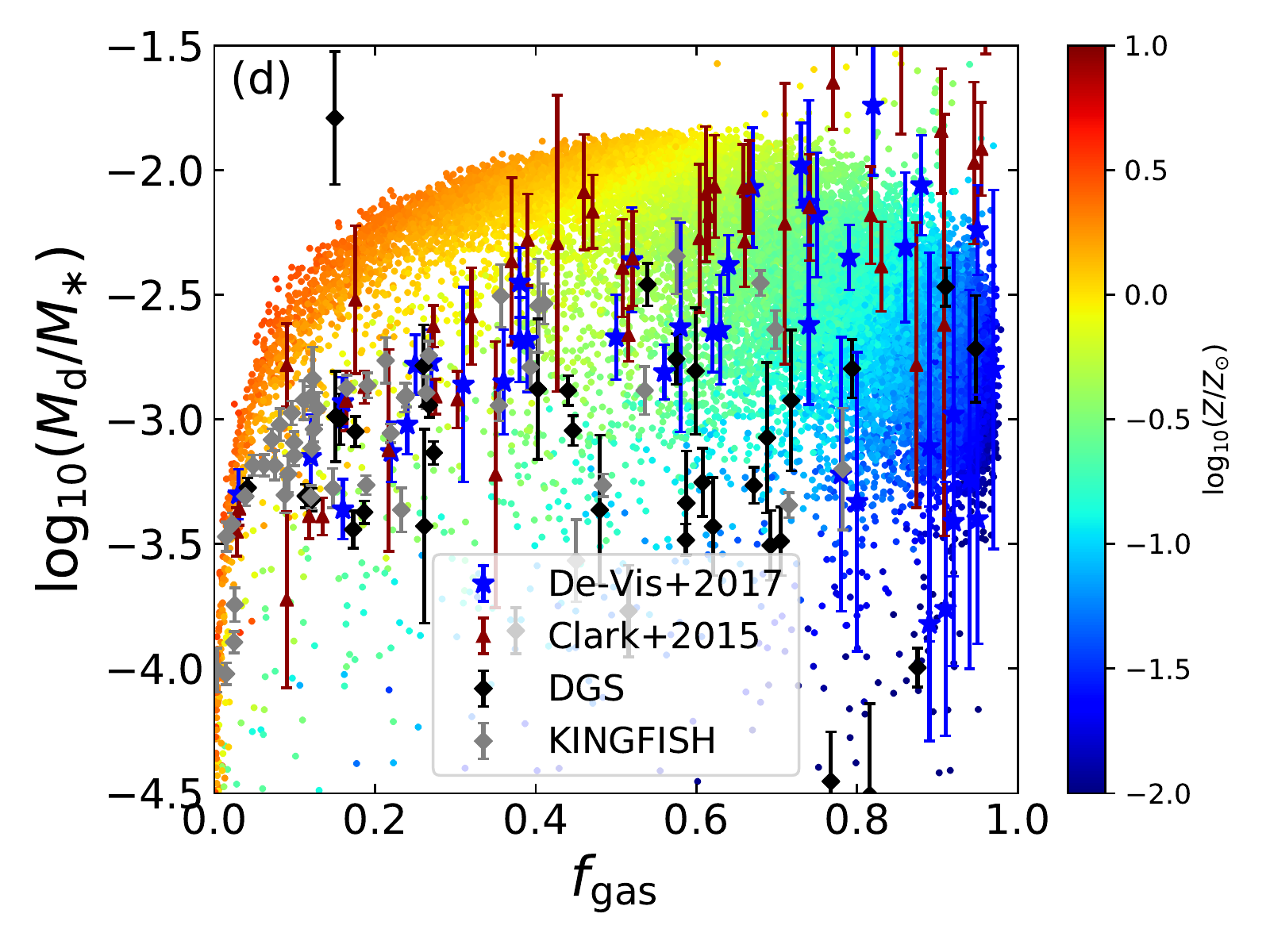}
	\caption{Scaling relations of dust-to-stellar mass ratio at $z = 0$
	against (a) metallicity, (b) stellar mass,
	(c) sSFR, and (d) gas fraction.
	Each point represents a galaxy with its colour indicating the stellar mass
	(panel (a)) or metallicity (panels (b), (c), and (d)) as shown in the colour bars.
	Observational data are from \citet{Remy-Ruyer:2014aa} (DGS and KINGFISH),
	\citet{Clark:2015aa}, and \citet{De-Vis:2017aa} as shown in the legend.
	}
	\label{Fig:D2S}
	\end{center}
\end{figure*}

In Fig.~\ref{Fig:D2S}a, we plot the $M_{\rm d}/M_{\ast}$--$Z$ relation.
The dust-to-stellar mass ratio is the highest at $Z \sim 1~ Z_\odot$.
The increase at sub-solar metallicities is driven by dust growth via accretion:
if dust is purely produced by stars, the dust-to-stellar mass ratio stays almost
constant at a level determined by the stellar dust yield.
At $Z \gtrsim 1~Z_\odot$, $M_{\rm d}/M_{\ast}$ drops, which is due to
the gas (and dust) consumption by star formation (i.e. astration).
Both DGS and KINGFISH datasets provide
information of dust mass, stellar mass and metallicity
and are overplotted for comparison.
The dust-to-stellar mass ratios in our simulation are broadly within the scatter of
the observational data at $Z\lesssim 0.3$ Z$_{\sun}$, although the observational samples
lack higher-metallicity galaxies for comparison.

In Fig.~\ref{Fig:D2S}b, we show
the $M_{\rm d}/M_{\ast}$--$M_{\ast}$ relation.
We observe that $M_{\rm d}/M_{\ast}$ has a peak
at $M_{\ast}\sim 10^{10}$\,$\Msun$ and
declines towards both the high-mass and the low-mass sides.
Because there is a strong correlation between $Z$ and $M_{\ast}$,
the rising trend at $M_{\ast}\lesssim 10^{10}$\,$\Msun$
is interpreted as driven by dust growth
caused by the metallicity increase.
The decreasing trend at $M_{\ast} \gtrsim 10^{10}$\,$\Msun$
is due to astration.
For comparison, we adopt the same
observational data as shown in Section \ref{Result:D2G}.
At $M_{\ast} \gtrsim 10^{9.5}$\,$\Msun$, the galaxies in the simulation
have higher $M_{\rm d}/M_{\ast}$ compared with the observations.
The reason for this overproduction is related to the excess of $\mathcal{D}$
in Fig.\,\ref{Fig:D2G}b (see the discussion in Section~\ref{Result:D2G}); that is,
we either need to adopt a more sophisticated AGN feedback model or need to
consider a possible bias in the observational samples. Indeed, the observational
sample does not contain metal-rich objects as shown in Fig.\,\ref{Fig:D2S}a.
At $M_{\ast} \lesssim 10^{9.5}$\,$\Msun$,
the observational data show a large variation,
and our prediction fits a large part of the $M_{\rm d}/M_{\ast}$ data.
\citet{McKinnon:2017aa} showed a decreasing trend of
$M_{\rm d}/M_{\ast}$ from $M_{\ast} = 10^7$ to $10^{11}$\,$\Msun$.
The higher dust condensation efficiencies and less efficient grain growth
by accretion in their simulation compared to ours
explain the difference between their result and ours.
We also provide further discussion about this discrepancy at low $M_{\ast}$
in Section~\ref{CompareObservation}.

Fig.~\ref{Fig:D2S}c shows the relation between
$M_{\rm d}/M_{\ast}$ and sSFR.
Although the overall correlation is weak, there is a trend that
$M_{\rm d}/M_{\ast}$ increases with increasing sSFR
for galaxies with $Z \gtrsim 1~ Z_\odot$.
There is also a weak tendency that
$M_{\rm d}/M_{\ast}$ decreases with increasing sSFR
for $Z \lesssim 1~ Z_\odot$.
The coverage of the observational data in this
$M_\mathrm{d}/M_\ast$--sSFR diagram is consistent with the
area covered by the simulation data.
\citet{Calura:2017aa} proposed, using their chemical evolution model,
that $M_{\rm d}/M_{\ast}$ strongly depends on the star formation history:
galaxies with a constant SFR
tend to have a flat $M_{\rm d}/M_{\ast}$ across all $M_{\ast}$,
while the starburst galaxies increase
$M_{\rm d}/M_{\ast}$ rapidly to the maximum level
in the early stage, and decrease it afterward.
In the latter case, a large dispersion
in $M_\mathrm{d}/M_\ast$ is expected.
Thus, the large dispersion in $M_\mathrm{d}/M_\ast$ at large
sSFR in Fig.~\ref{Fig:D2S}c could be a natural consequence
of their high star formation activities.
The observational sample adopted in Fig.~\ref{Fig:D2S}c
contains starburst galaxies and indeed
shows a large dispersion in $M_\mathrm{d}/M_\ast$.

We show the $M_{\rm d}/M_{\ast}$--$f_{\rm gas}$ relation
in Fig.~\ref{Fig:D2S}d.
In the simulation $M_{\rm d}/M_{\ast}$ has a peak around $f_{\rm gas} \sim 0.6$
with lower values on both high and low $f_{\rm gas}$ sides.
The low $M_{\rm d}/M_{\ast}$ at low $f_{\rm gas}$ end is interpreted as
an early phase of dust enrichment in gas-rich galaxies.
The dust-to-stellar mass ratio
drops at $f_{\rm gas}\lesssim 0.2$
because star formation has consumed a large fraction of gas and dust
in high-metallicity galaxies by $z=0$.
Our simulation reproduces the observational trend in the
$M_{\rm d}/M_{\ast}$--$f_{\rm gas}$ relation correctly, and it also
covers the large dispersion at high $f_\mathrm{gas}$.
There is a slight excess of $M_{\rm d}/M_{\ast}$
at $f_{\rm gas}$ $\lesssim 0.4$ relative to the observational data,
which is associated with the overproduction of $M_{\rm d}/M_{\ast}$ around
$M_\ast\sim 10^{10}$ M$_{\sun}$
(see the discussion for the excess of $M_{\rm d}/M_{\ast}$ above).


\begin{figure*}
	\begin{center}
	\includegraphics[width=0.475\textwidth]{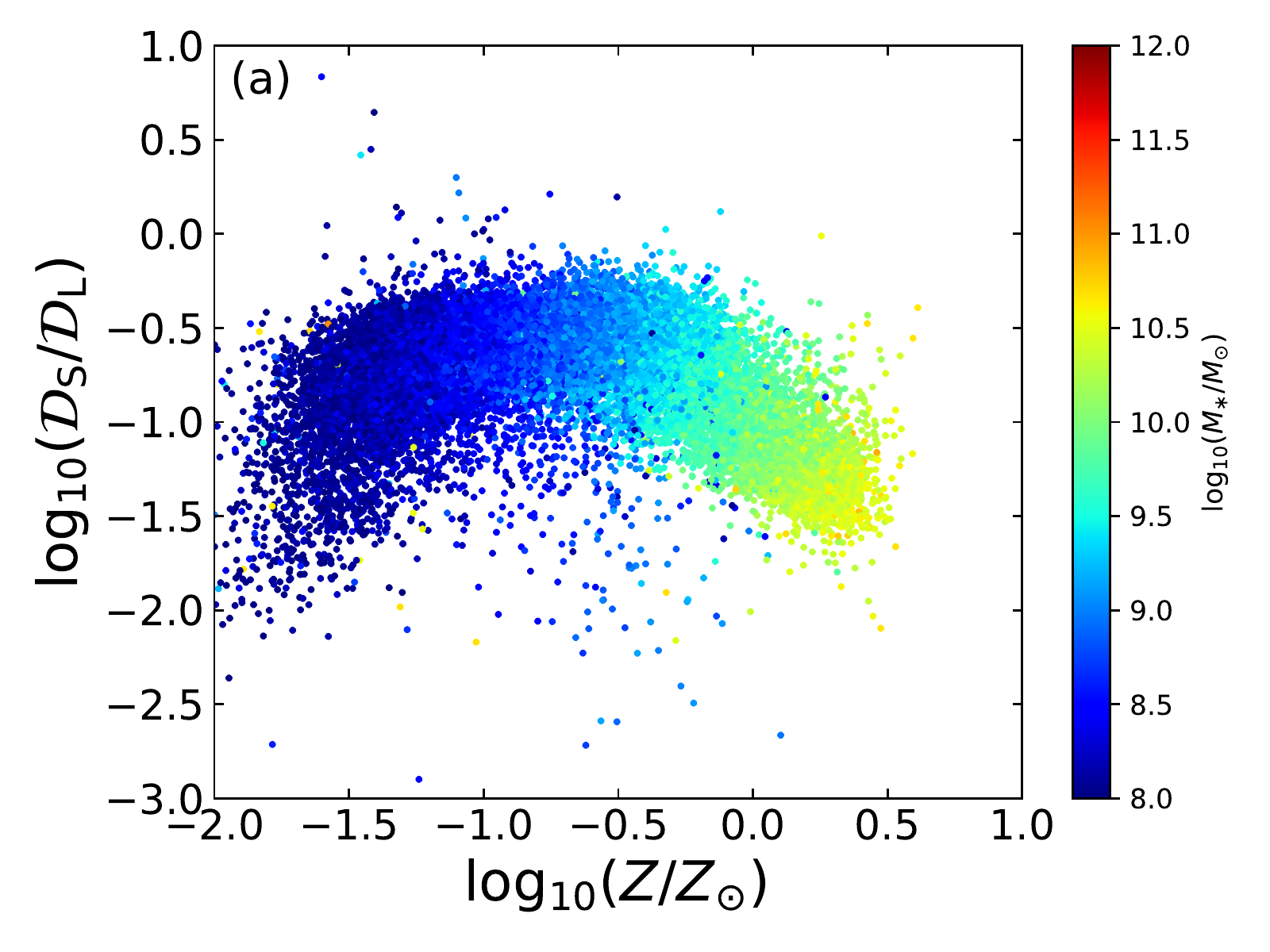}
	\includegraphics[width=0.475\textwidth]{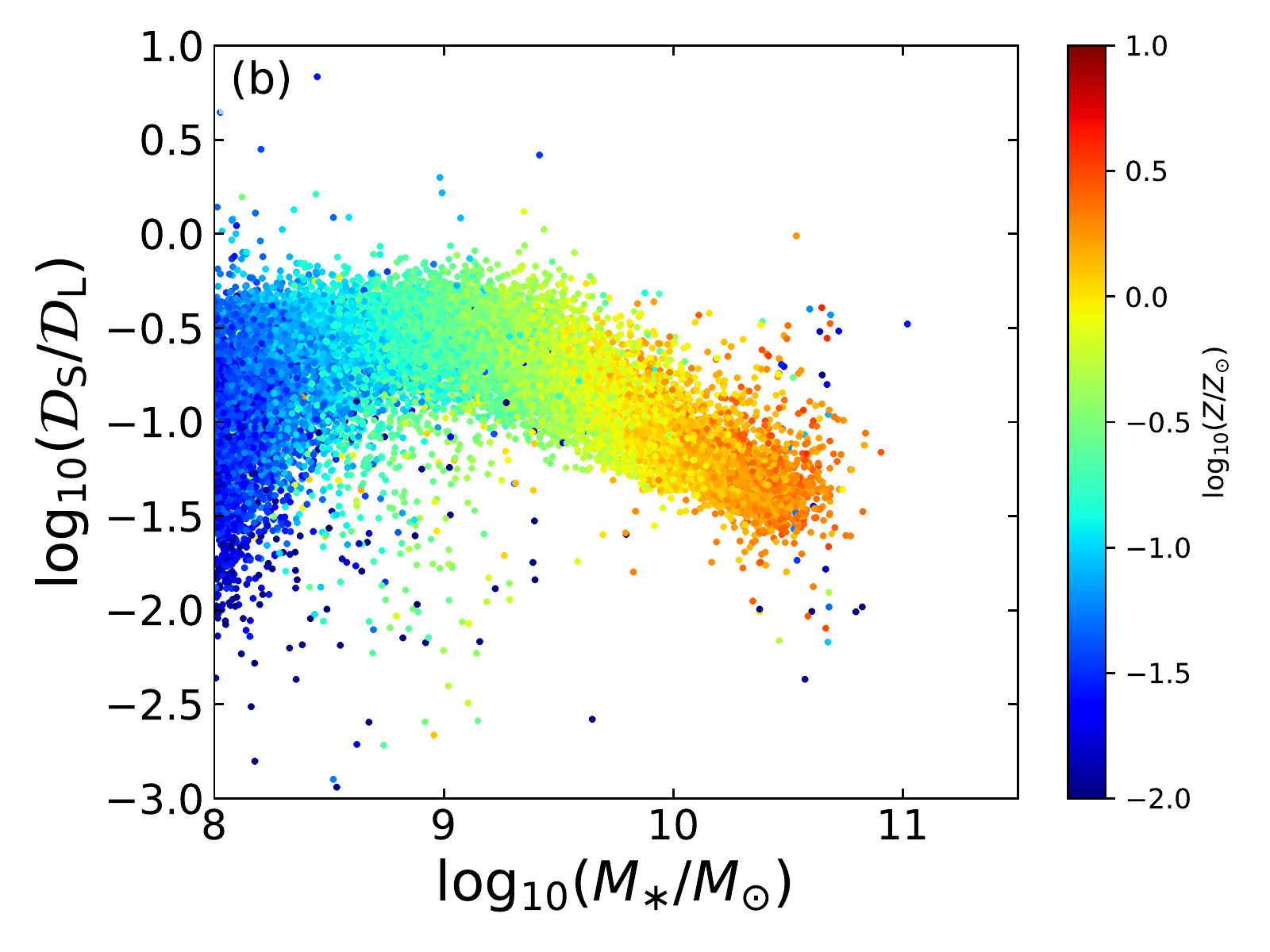}
	\includegraphics[width=0.475\textwidth]{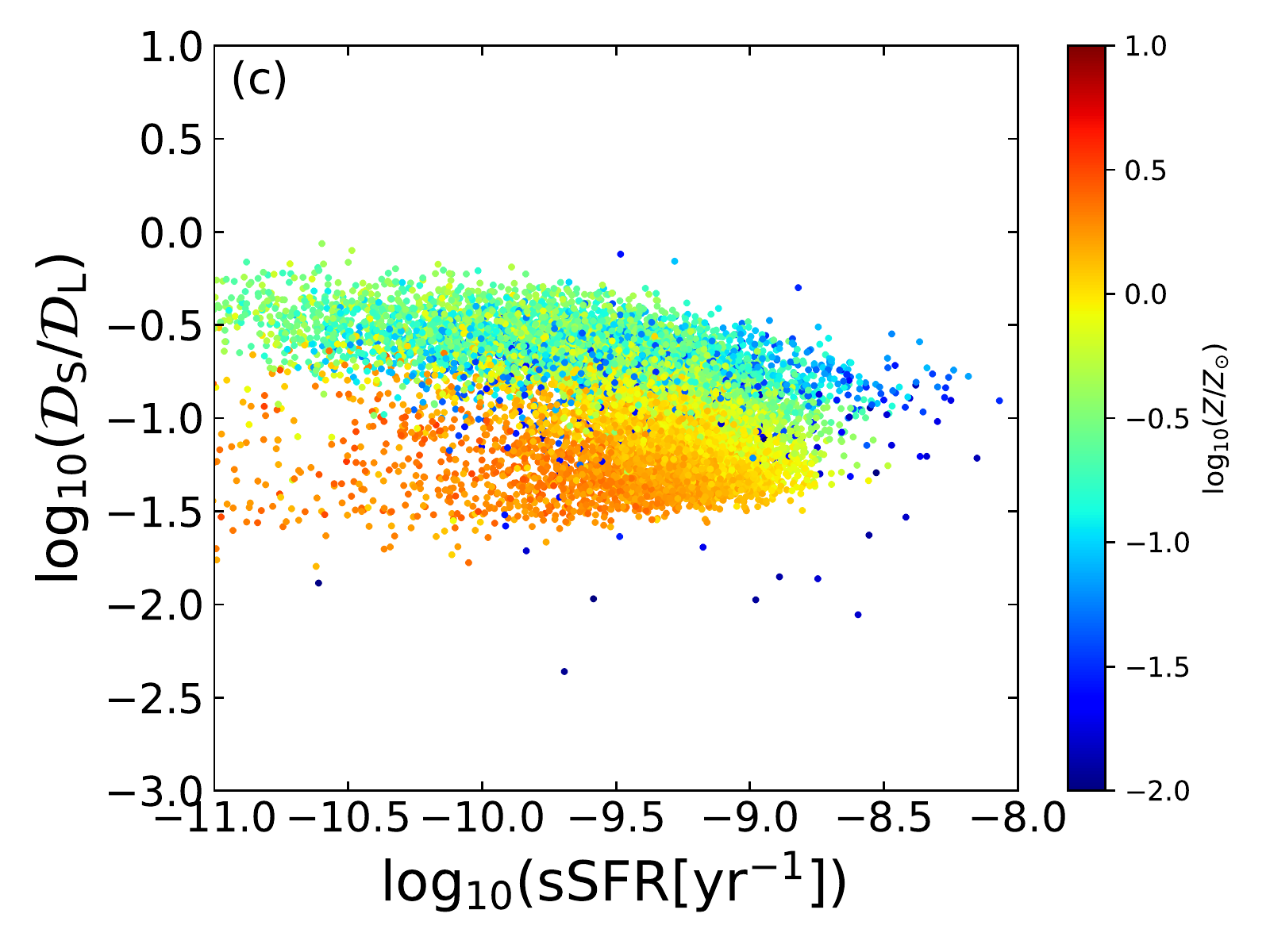}
	\includegraphics[width=0.475\textwidth]{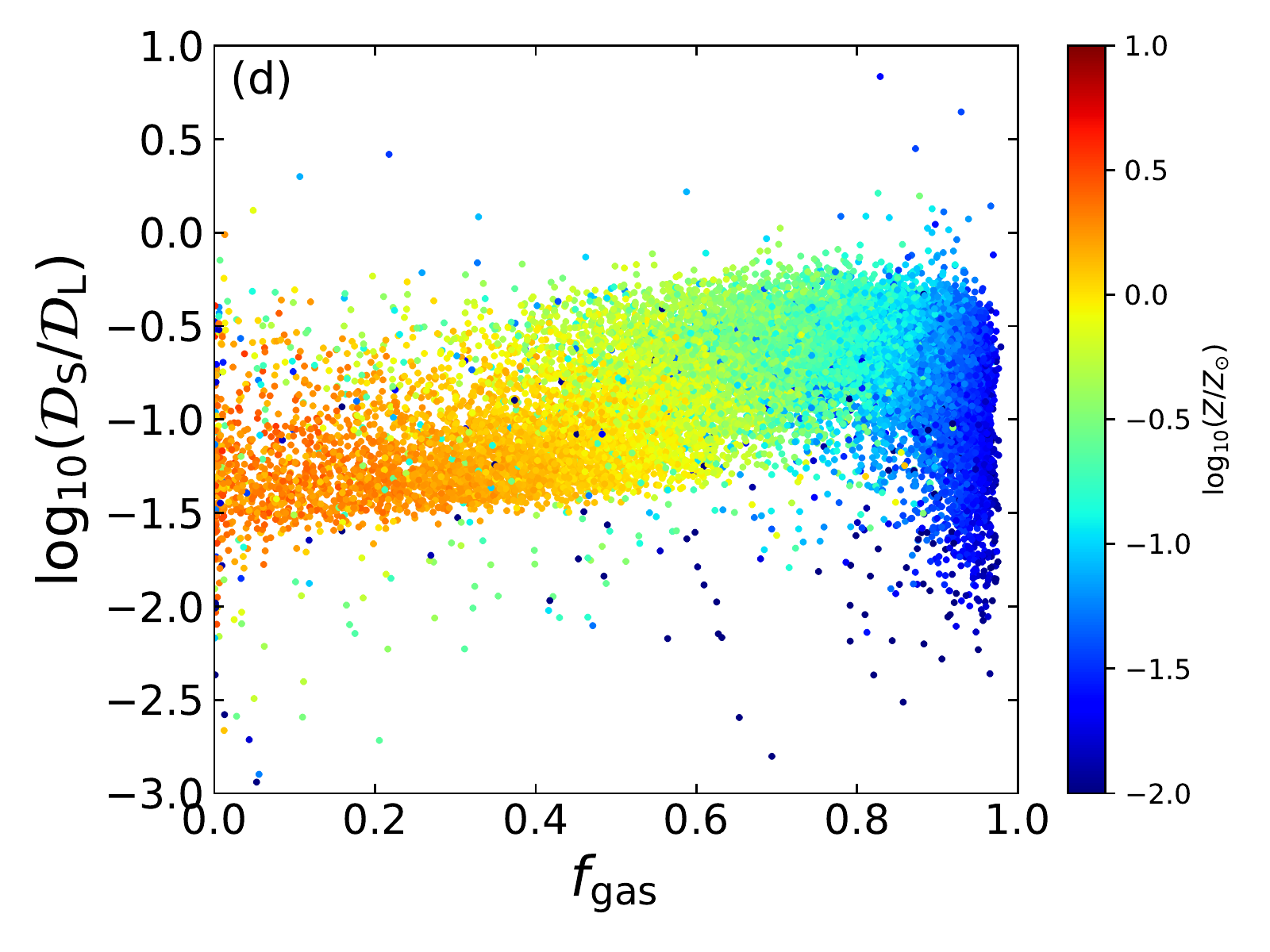}
	\caption{Scaling relations of the small-to-large grain abundance ratio at $z = 0$
	against (a) metallicity, (b) stellar mass,
	(c) sSFR, and (d) gas fraction.
	Each point represents a galaxy with its colour indicating the stellar mass
	(in panel a) or metallicity (panels b, c, and d) as shown in the colour bars. }
	\label{Fig:S2L_1}
	\end{center}
\end{figure*}%

\subsection{Small-to-large grain abundance ratio}

Our dust enrichment treatment provides the information
on grain size distribution represented by
the abundances of large and small grains.
Here we examine the behaviour of the small-to-large grain abundance ratio,
$\stol$, in terms of $Z$, $M_{\ast}$, sSFR and $f_{\rm gas}$.
In our simulation, the production of small grains is governed by shattering and accretion,
while the increase of large grains is dominated by stellar dust production and coagulation.
By investigating $\stol$, we are able to understand how those processes affect the dust
evolution.
Because there is no statistical, observational data for grain size distribution,
we only describe our theoretical predictions here. The effect of
grain size distribution on extinction curves is discussed later in
Section~\ref{Result:extinction}.

Fig.~\ref{Fig:S2L_1}a shows the $\stol$--$Z$ relation.
The small-to-large grain abundance ratio increases
between $Z \sim 0.01~ Z_\odot$ and $\sim 0.1~ Z_\odot$,
remains almost constant at $0.1 \lesssim Z \lesssim 0.3~ Z_\odot$,
and declines at $Z \gtrsim 0.3~ Z_\odot$.
The dust production in low-metallicity galaxies
is dominated by stellar dust production.
Shattering is the source of small grains in this phase.
Accretion, which only works on small grains,
efficiently raises $\stol$ when the small grain abundance
and the metallicity are high enough.
Thus, $\stol$ increases by more than an
order of magnitude between
$Z \sim 0.01~ Z_\odot$ and $Z \sim 0.1~ Z_\odot$.
At $0.1 \lesssim Z \lesssim 0.3~ Z_\odot$,
coagulation becomes efficient. In this phase,
the increase of small grain by accretion and shattering and
the increase of large grain by coagulation
are comparable, so that $\stol$ shows a flat trend.
At $Z \gtrsim 0.3~ Z_\odot$, coagulation
is stronger than accretion and shattering,
so that $\stol$ decreases with increasing $Z$.

The $\stol$--$M_{\ast}$ relation, shown in Fig.~\ref{Fig:S2L_1}b, is
similar to the $\stol$--$Z$ relation because of the tight correlation
between stellar mass and metallicity.
There is a huge dispersion in $\stol$ between
$M_\ast\sim 10^8$ and $10^{8.5}$\,$\Msun$, which is created by the
increase of $\mathcal{D}_{\rm s}$ as a result of accretion.
There is a plateau between $M_\ast \sim 10^{8.5}$ and $10^{9}\,\Msun$
caused by the balance of accretion, shattering and coagulation.
At $10^{9} \lesssim M_{\ast} \lesssim 10^{11}$\,$\Msun$,
$\stol$ decreases with increasing $M_{\ast}$, because
coagulation reduces small grains and increase large grains.

Fig.~\ref{Fig:S2L_1}c presents the relation between $\stol$ and
sSFR. We observe
a weak tendency that galaxies with higher sSFR
have lower $\stol$. This is because sSFR tends to be lower in metal-rich
(or gas-poor) objects.

Fig.~\ref{Fig:S2L_1}d shows the relation between $\stol$ and $f_{\rm gas}$.
Galaxies with $f_{\rm gas} \gtrsim$ 0.8 have a wide range of $\stol$, whose
variety is driven by accretion.
Between $f_{\rm gas} = 0.6$ and $0.8$,
$\stol$ remains roughly constant since coagulation counterbalances
shattering and accretion.
$\stol$ decreases from $f_{\rm gas}$ = 0.6 to 0
because coagulation is stronger than other processes;
here, we recall that galaxies with low $f_\mathrm{gas}$
tend to have high dust-to-gas ratio, which is a favourable condition for coagulation.

In summary, the evolution of grain size distribution is roughly understood
as follows.
Stellar dust production and shattering dominate the early stage of
dust evolution when a galaxy is small, metal-poor and
extremely gas-rich. In this phase, the grain abundance is dominated by large
grains.
In turn, accretion becomes the dominant driver of evolution at
$Z \sim 0.03$--0.3 Z$_\odot$,
$M_{\ast} \sim 10^8 - 10^{8.5}$\,$\Msun$ and $f_{\rm gas} \gtrsim 0.8$,
and accretion drastically increases the abundance of small grains.
There is a phase in which galaxies have roughly constant $\stol$ because
the formation of small grains by accretion and shattering
are compensated by large grain creation by coagulation. This phase
corresponds to
$Z \sim 0.1-0.3\,Z_\odot$, $M_{\ast} \sim 10^{8.5}-10^{9}$\,$\Msun$ and
$f_{\rm gas} \sim 0.8-0.6$.
In more evolved galaxies with $Z \gtrsim 0.3~ Z_\odot$,
which typically have $M_{\ast} \gtrsim 10^{9.5}$\,$\Msun$
and $f_{\rm gas}$ $\lesssim 0.6$, coagulation is somewhat stronger
than shattering and accretion, so that $\stol$ declines.
At this stage, the processes in the ISM are much more efficient than the stellar dust production
in determining both the grain size distribution and the total dust abundance.


\subsection{Redshift evolution}
\label{RedshiftEvolution}

Our simulation allows us to examine the evolution up to $z \sim 5$.
Above $z \sim 5$, the simulation did not
produce enough galaxies with $M_\ast \gtrsim 10^8$\,$\Msun$
due to the limited  simulation volume.
It requires a larger simulation volume 
for sufficient statistical data, but the
computational cost will be much higher if the same spatial and mass
resolution is required.

\begin{figure}
	\begin{center}
	\includegraphics[width=0.475\textwidth]{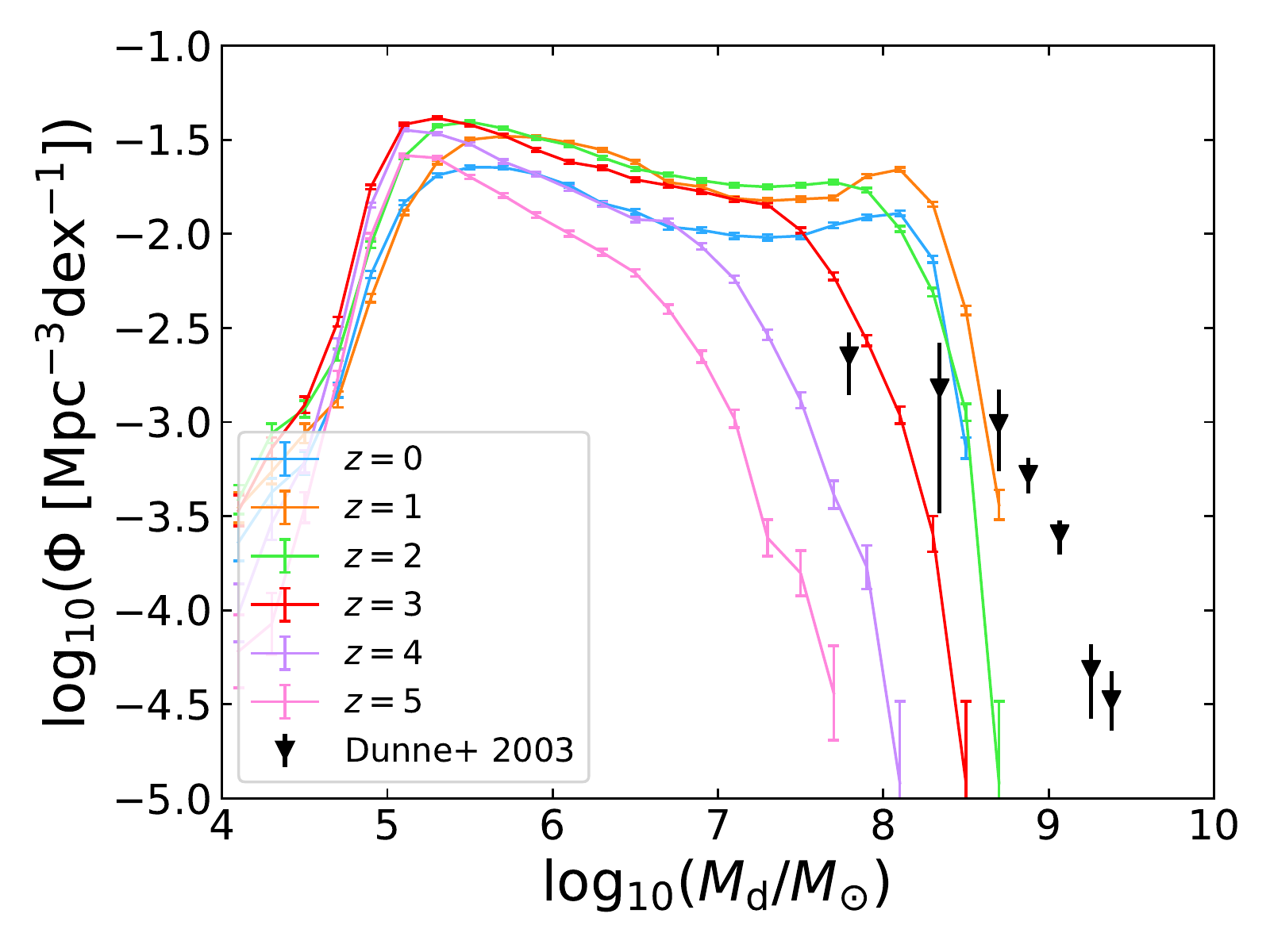}
	\includegraphics[width=0.475\textwidth]{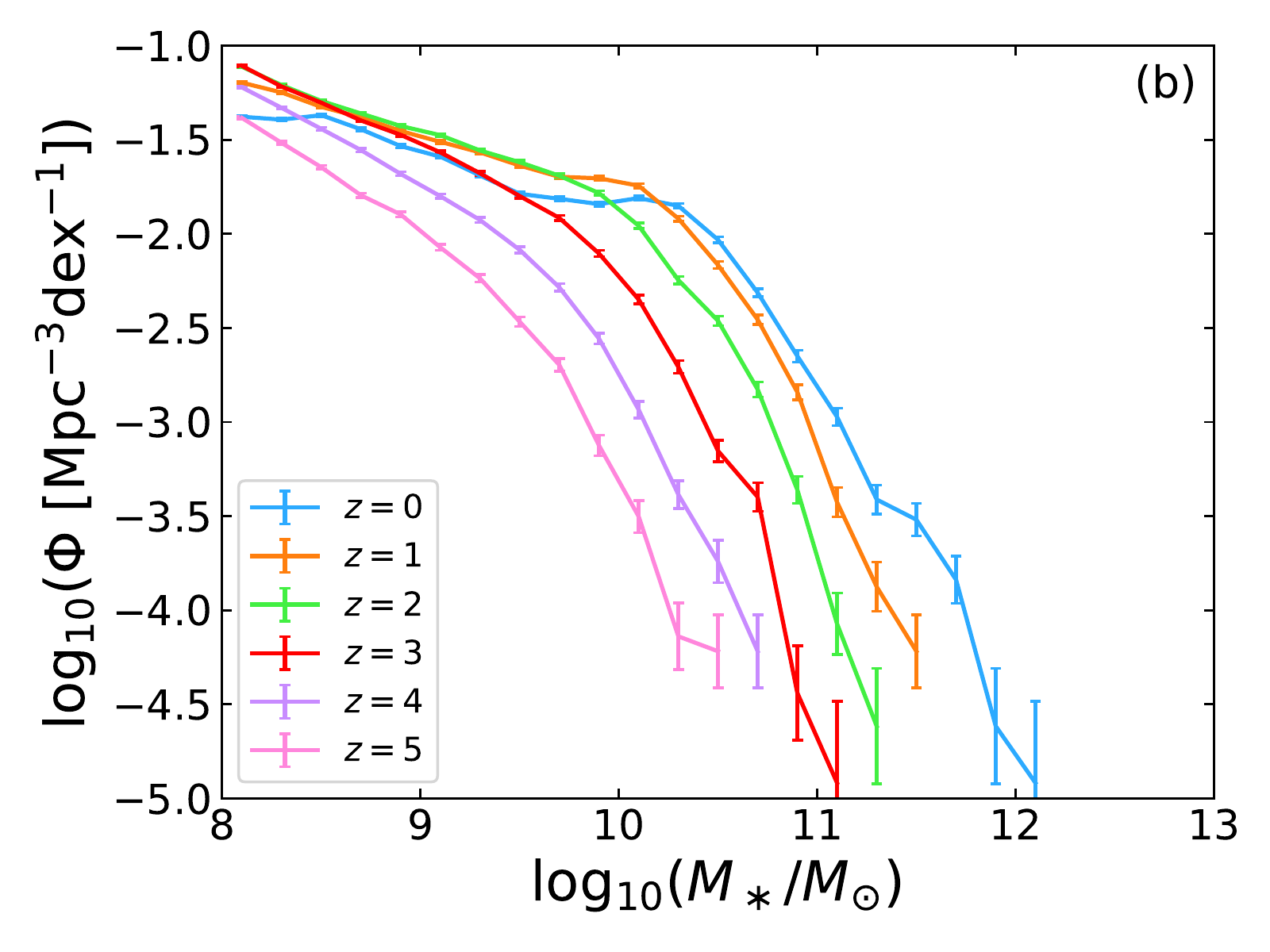}
	\caption{Redshift evolution of (a) dust mass function and
	(b) stellar mass function.
	The redshifts are shown in the legend.
	Black points are the observed dust mass
	function at $z \sim 2.5$ from \citet{Dunne:2003aa}.
	The error bars are based on the Poisson statistics.
	}
	\label{Fig:MF_dust_z}
	\end{center}
\end{figure}

In Fig.~\ref{Fig:MF_dust_z}, we show the redshift evolution of
dust mass function.
We also show the galaxy stellar mass functions in the bottom panel for reference.
The decrease of the dust mass function at $M_{\rm d} \lesssim 10^{5}$\,$\Msun$
is due to the limited mass resolution in the simulation.
Recall that we select galaxy with $M_{\ast} > 10^{8}$\,$\Msun$.
Between $M_{\rm d} = 10^{5}$ and $10^{8}$\,$\Msun$,
at a fixed dust mass,
the galaxy number density increases from $z=5$ to 2,
remains almost constant between $z=2$ and 1
and decreases at $z \lesssim 1$.
These variations from $z=5$ to 0 roughly trace the evolution of
the comoving dust mass density in the Universe
\citep[Paper I;][]{Driver:2018aa}.

We find that, unlike the high-mass end of the stellar mass function,
the dust mass function does not have an extended tail at the massive end.
It seems that the dust growth is limited to
$M_{\rm d} \lesssim 10^9$\,$\Msun$.
This upper dust-mass limit is caused by astration (i.e.\ consumption of gas and dust into stars),
which is more significant than the dust formation
in massive galaxies with $M_{\rm d} \gtrsim 10^8$\,$\Msun$ at $z \lesssim 2$;
therefore the galaxy number density decreases at the high dust-mass end
from $z \sim~ 1$ to 0.

Accretion quickly raises the dust abundance and establishes dust-rich galaxies
with $M_{\rm d} \gtrsim 10^8$\,$\Msun$ at $z \lesssim 2$.
On the other hand, dust-rich galaxies also suffer astration, and their dust mass is
limited to $M_{\rm d} \lesssim 10^9$\,$\Msun$ as discussed above.
Therefore, the interplay between accretion and astration creates
a bump at $M_{\rm d} \sim 10^{8.2}$\,$\Msun$  at $z = 1$ and 0 in our model.
Besides, as discussed in Section \ref{result:DMF},
our simple AGN feedback model somewhat fails to
reduce the gas mass by outflows, which makes some galaxies
overabundant in dust mass, even though the dust-to-gas ratios are not overpredicted.

Overall, the dust abundance is the highest between $z$ = 2 and 1 in our
simulation.
In contrast, \citet{McKinnon:2017aa} and \citet{Popping:2017aa}
predicted that the dust mass function increases from $z$ = 2 to 0
at the high-mass end.
We show the observed dust mass function at $z \sim~ 2.5$ by \citet{Dunne:2003aa},
which, compared with our simulation, has a higher galaxy number density
at $M_\mathrm{d}\gtrsim 10^9$ M$_{\sun}$,
and a lower galaxy number density at lower dust masses.
The caveat is that their results are based on the SCUBA surveys
with a large beam size ($\sim 30 \arcsec$).
Such a low spatial resolution tends to blend multiple sources
in a beam, which could cause an overestimate of the number of high dust-mass galaxies,
and could miss faint galaxies around or below the confusion limit
\citep{Karim:2013aa,WangWH:2017aa}.

\begin{figure}
	\begin{center}
	\includegraphics[width=0.475\textwidth]{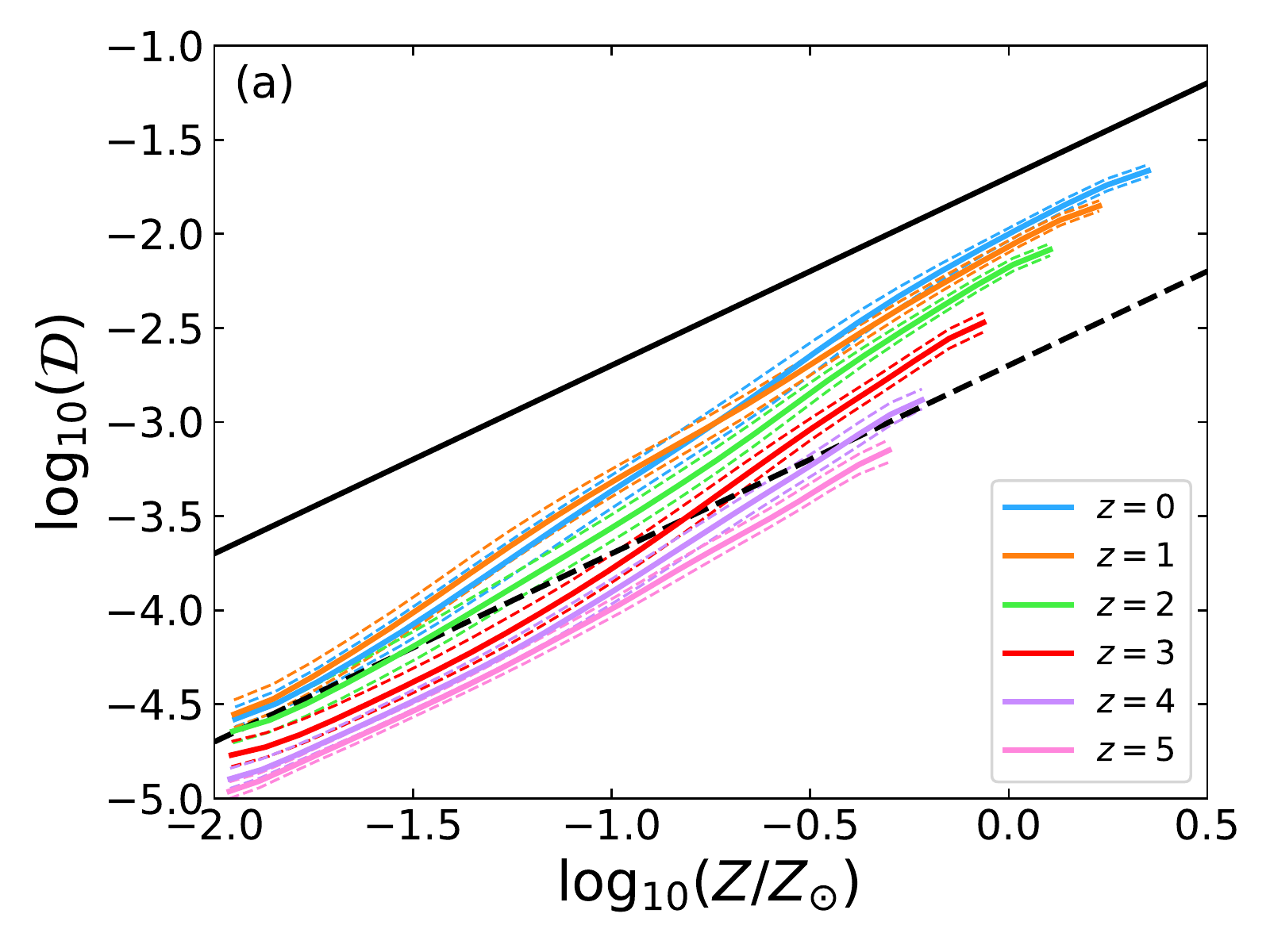}
	\includegraphics[width=0.475\textwidth]{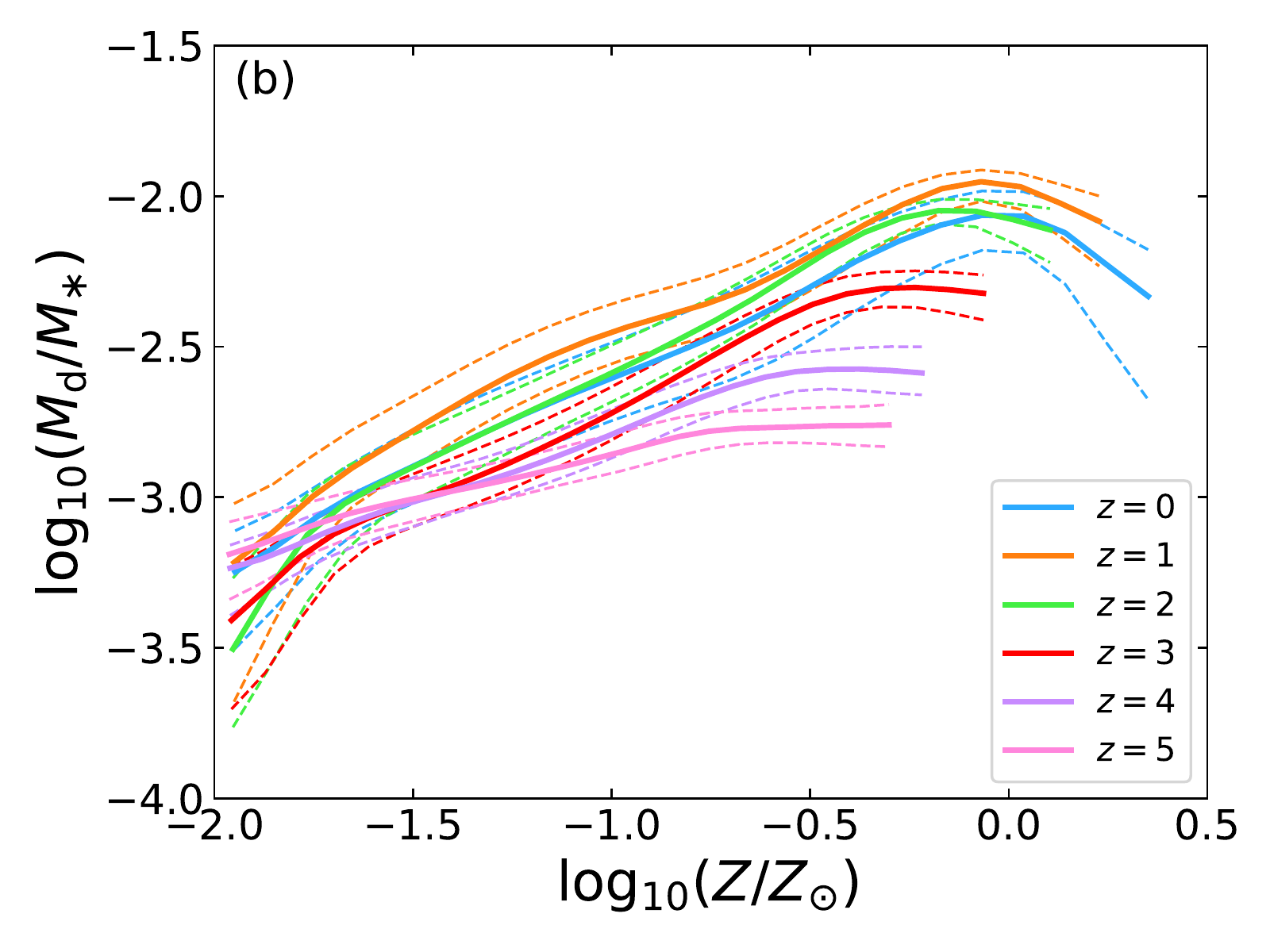}
	\includegraphics[width=0.475\textwidth]{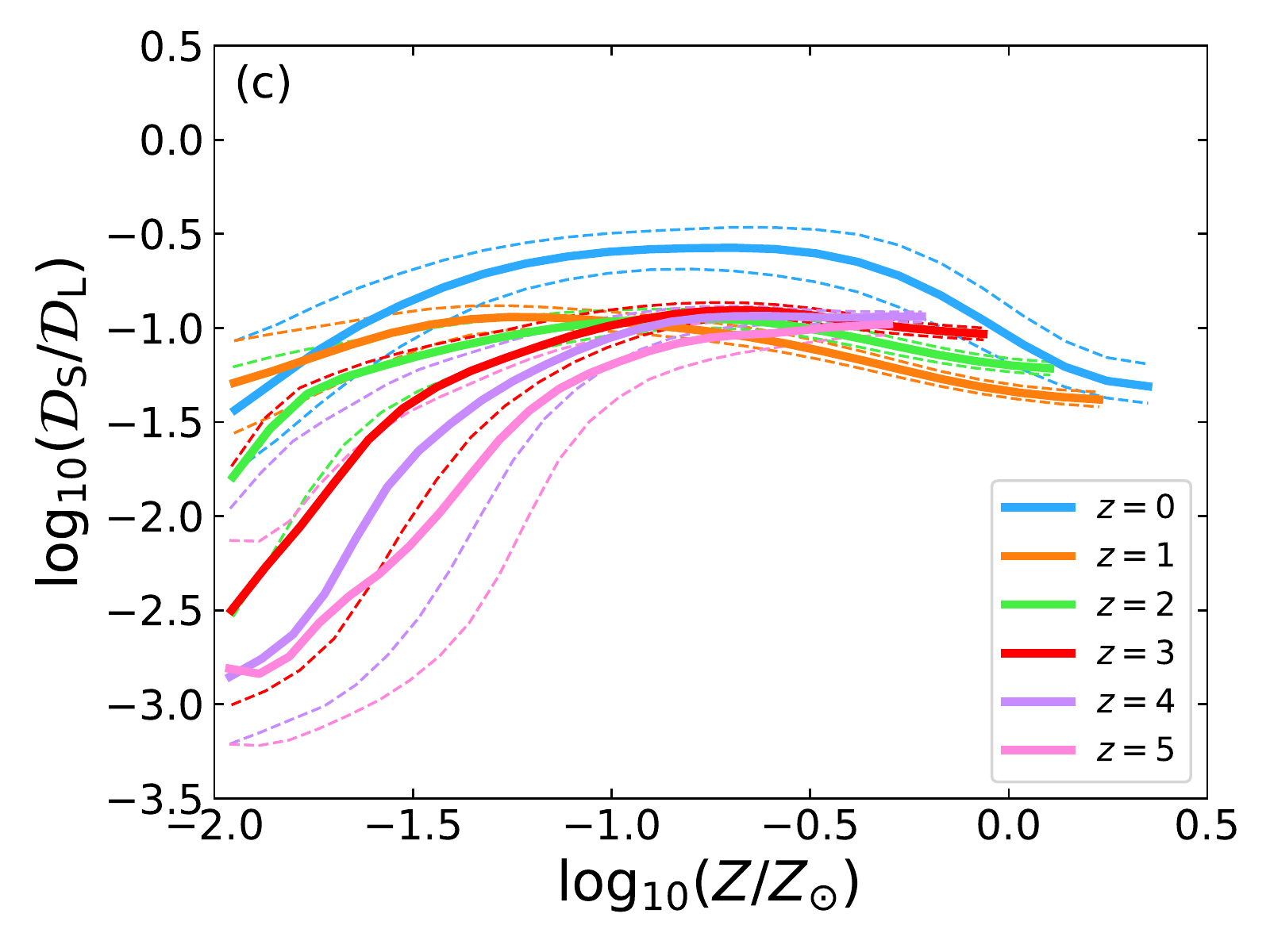}
	\caption{Redshift evolution of the ({a}) $\mathcal{D}-Z$,
	({b}) $M_{\rm d}/M_{\ast}$--$Z$ and ({c}) $\stol$--$Z$ relations.
	The thick solid lines are the median in each metallicity bin at each redshift.
	The thin dashed lines are 25th and 75th percentiles in each bin.
	The blue, orange, green, red, purple and pink lines correspond to $z =0$, 1, 2, 3, 4, and  $5$, respectively.
	}
	\label{Fig:redshift}
	\end{center}
\end{figure}

Next, we examine how the basic scaling relations
evolve with cosmic time.  Since dust enrichment is
strongly linked to metal enrichment, it is interesting to show the
relations between dust abundance indicators ($\mathcal{D}$ and
$M_\mathrm{d}/M_\ast$) and metallicity. We also examine
the redshift evolution of $\mathcal{D}_\mathrm{S}/\mathcal{D}_\mathrm{L}$.
In Fig.~\ref{Fig:redshift}, we show the $\mathcal{D}$--$Z$ and
$M_{\rm d}/M_{\ast}$--$Z$ relations from $z =0$ to 5.
We derive the median of $\mathcal{D}$
and $M_{\rm d}/M_{\ast}$ in every logarithmic metallicity bin
smoothed by a gaussian kernel along the metallicity axis with
a standard deviation $\sigma = 0.1$ dex to avoid the statistical
fluctuations.

In Fig.~\ref{Fig:redshift}a,
we present the redshift evolution of $\mathcal{D}$--$Z$ relation.
At
a fixed metallicity,
$\mathcal{D}$ tends to increase with decreasing redshift.
At $z \gtrsim 4$, most galaxies have $\mathcal{D}$ lower than the values
expected from stellar dust production
($f_\mathrm{in}Z$; dashed line in Fig.~\ref{Fig:redshift}a).
Active star formation in high-redshift galaxies
makes SN dust destruction efficient; thus $\mathcal{D}$ is suppressed.
This also happens in low metallicity galaxies at $z \sim 3$.
Although there are a small number of galaxies whose metallicity
is nearly solar at $z$ = 5,
their dust-to-gas ratios remain low compared with lower redshifts.
As mentioned in Section \ref{Result:D2G},
the nonlinear increase of $\mathcal{D}$ above
a certain metallicity is due to dust growth by accretion.
In the following, we refer to the metallicity at which accretion starts
to dominate the increase of $\mathcal{D}$ as the `turning point'.

From Fig.~\ref{Fig:redshift}a, it is clear that the turning point
shifts towards higher metallicity with increasing redshift at $z \gtrsim 1$.
\citet{Asano:2013ab} pointed out that the turning point
(referred to as the `critical metallicity' in their paper)
is determined by the ratio between the star formation time-scale and
the dust growth time-scale.
A shorter star-formation time-scale caused by the rich gas content in
high-redshift galaxies pushes the turning point to a higher metallicity at higher redshift.

In the $M_{\rm d}/M_{\ast}$--$Z$ relation shown in Fig.~\ref{Fig:redshift}b,
the value of $M_{\rm d}/M_{\ast}$ increases from $z \sim 5$ to 1 and
decreases from $z \sim 1$ to 0 at a fixed metallicity
except at $Z \sim 0.01 Z_\odot$.
At $z \sim 5$,  dust growth by accretion is not efficient yet.
At lower redshifts, $M_{\rm d}/M_{\ast}$ has a rapid increase
above the turning-point metallicity.
As we have already observed for $\mathcal{D}$ above,
the turning point shifts to lower metallicities as
the redshift decreases.
We find that the metallicity at which $M_{\rm d}/M_{\ast}$ peaks
increases from $z \sim 5$ to 1
and remains at a similar level between $z \sim 1$ and 0.
The decrease of $M_{\rm d}/M_{\ast}$ at high metallicity is due to astration.
The maximum value of $M_{\rm d}/M_{\ast}$ rises from $z=5$ to 1,
and decreases at $z \lesssim 1$.
This complex behaviour is due to the competition between the monotonic
increase of $\mathcal{D}$ along the redshift and the monotonic decrease of
$M_\mathrm{gas}/M_\ast$ (note that
$M_\mathrm{d}/M_\ast =\mathcal{D} \times M_\mathrm{gas}/M_\ast$).

Finally, we examine the redshift evolution of
grain size distribution in Fig.\ \ref{Fig:redshift}c.
The evolution of the $\stol$--$Z$ relation is
much more complicated than the above two relations because there are
two more processes directly affecting the grain size; namely, coagulation and shattering.
At $z \gtrsim 3$, a large portion of low-metallicity galaxies have
$\stol \lesssim 0.03$ since small grains are produced by only shattering
in the beginning.
\citet{Hirashita:2015aa} also showed that a shorter star formation time-scale
makes a lower $\stol$, because the enrichment of large grains by stellar dust
production proceeds before shattering
produces a significant amount of small grains.
More active star formation at higher redshifts is a natural
consequence of a rich gas content.  
After accretion becomes efficient,
$\stol$ rises to $\sim~10^{-1}$ quickly.
This $\stol$ value is regulated by the balance among shattering, coagulation
and accretion.
We observe a weak decline toward high metallicity.
This decline is caused by coagulation and is more prominent for lower-redshift galaxies,
which have higher dust-to-gas ratios. 
Galaxies at $z = 0$ have higher $\stol$ than those at other redshifts
due to lower dense-gas fractions (coagulation is suppressed and
shattering is enhanced) and longer star formation (= chemical enrichment)
time-scales (i.e.\
shattering produces more small grains while the chemical enrichment proceeds).


\subsection{Extinction curves}
\label{Result:extinction}

A viable way of testing the evolution of grain size distribution is to examine
the extinction curves \citep{Asano:2013ab}.
The steepness of extinction curve indicates the
small-to-large grain abundance ratio at least qualitatively.
Based on $\stol$ obtained above, we calculate the extinction curves in this
subsection for future tests, and also attempt to compare with some
observed extinction curves.

We calculate the extinction curve following the formulation in
Section~\ref{subsec:extinc_curve}.
For the dust species adopted in our model, the
2175\,\AA\ bump and far-ultraviolet (FUV) rise are caused by
small dust grains, so a higher $\stol$ leads to an extinction curve
with a stronger 2175\,\AA\ bump and a steeper FUV rise.
Although the strength of 2175\,\AA\ bump is not robust against the
change of grain species \citep{Hou:2016aa}, the steepness of FUV rise
is suitable for tracing the increase
of the small-grain abundance relative to the large-grain abundance.
Note also that
determining the grain size distribution from an observed extinction curve
is somewhat dependent on the assumed grain species
\citep{Zubko:1996aa}. In this sense, fixing the grain species is useful
for the first step
to isolate the trend caused by the evolution of grain size distribution.
The following discussions on the evolutionary behaviour of $\stol$ is
supported by the discussions in Section~\ref{RedshiftEvolution}.

\begin{figure*}
	\begin{center}
	\includegraphics[width=0.32\textwidth]{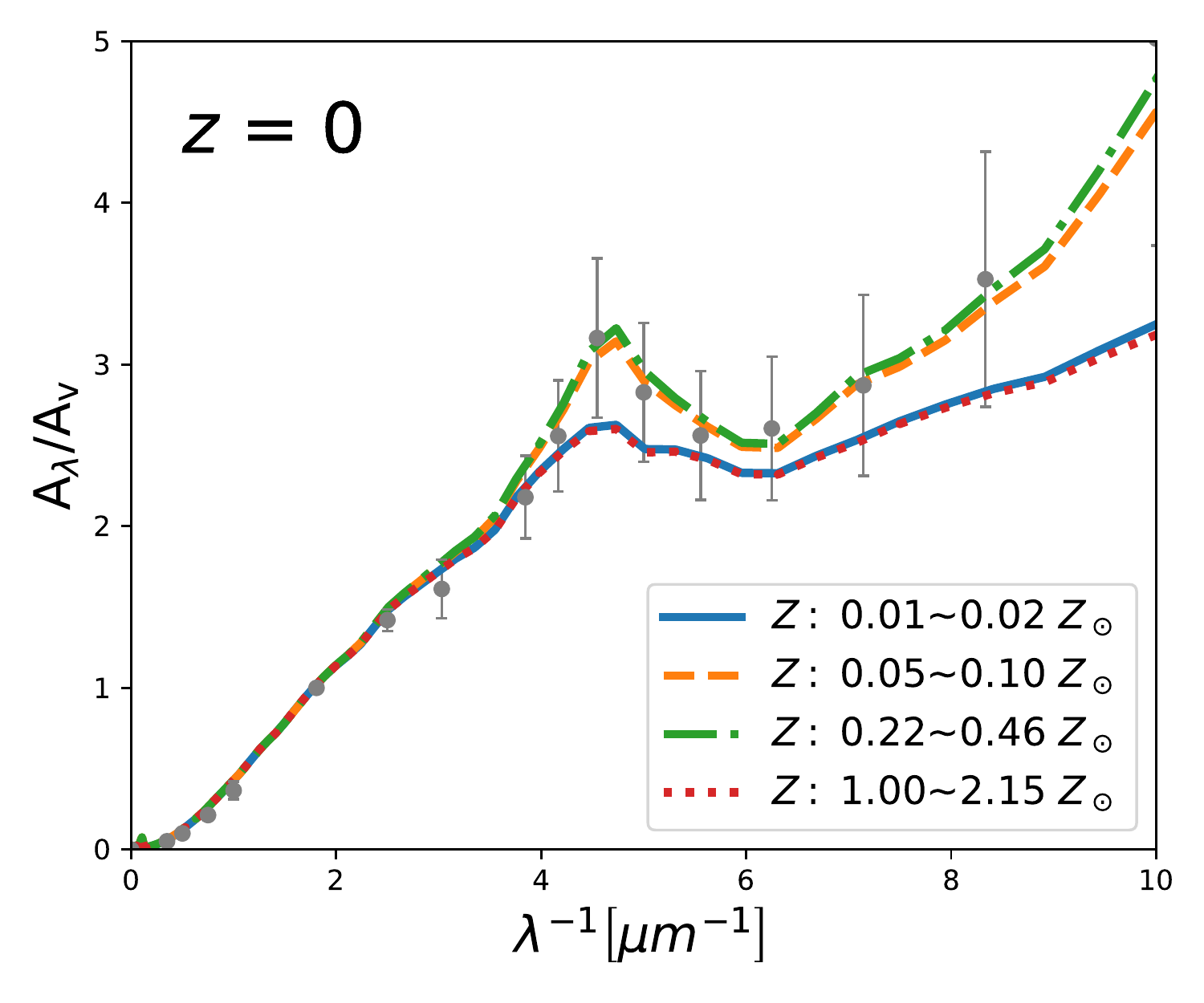}
	\includegraphics[width=0.32\textwidth]{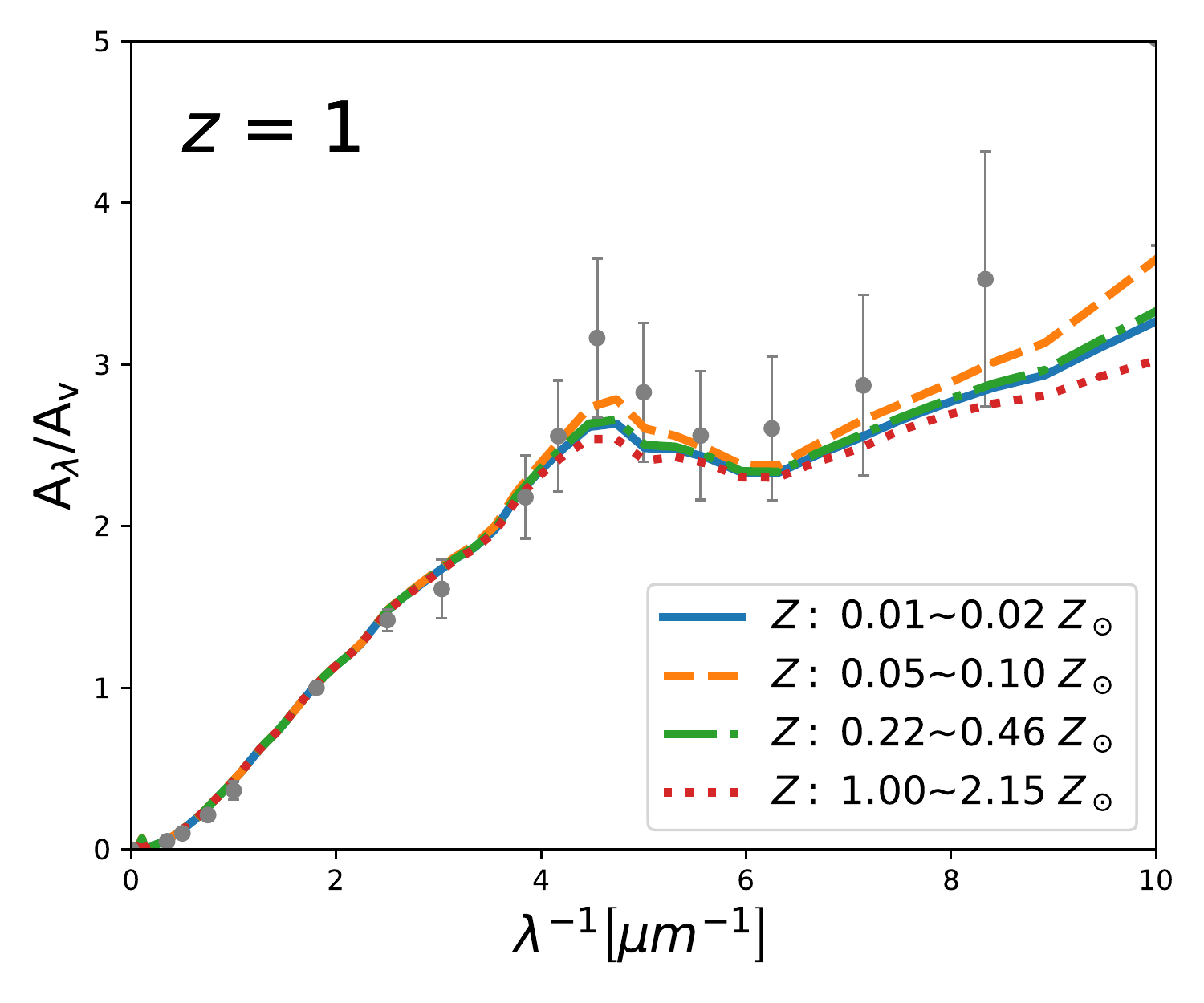}
	\includegraphics[width=0.32\textwidth]{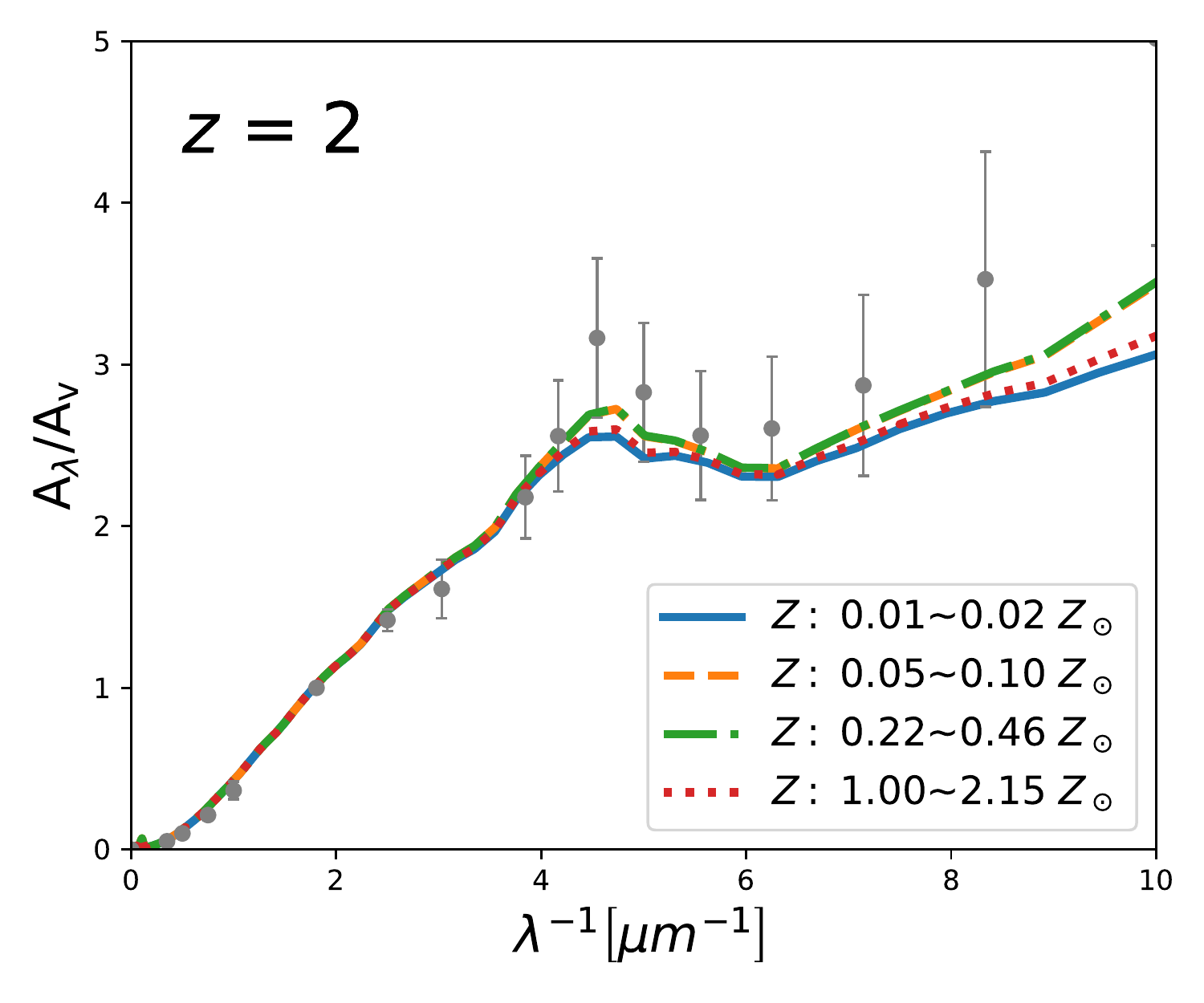}
	\includegraphics[width=0.32\textwidth]{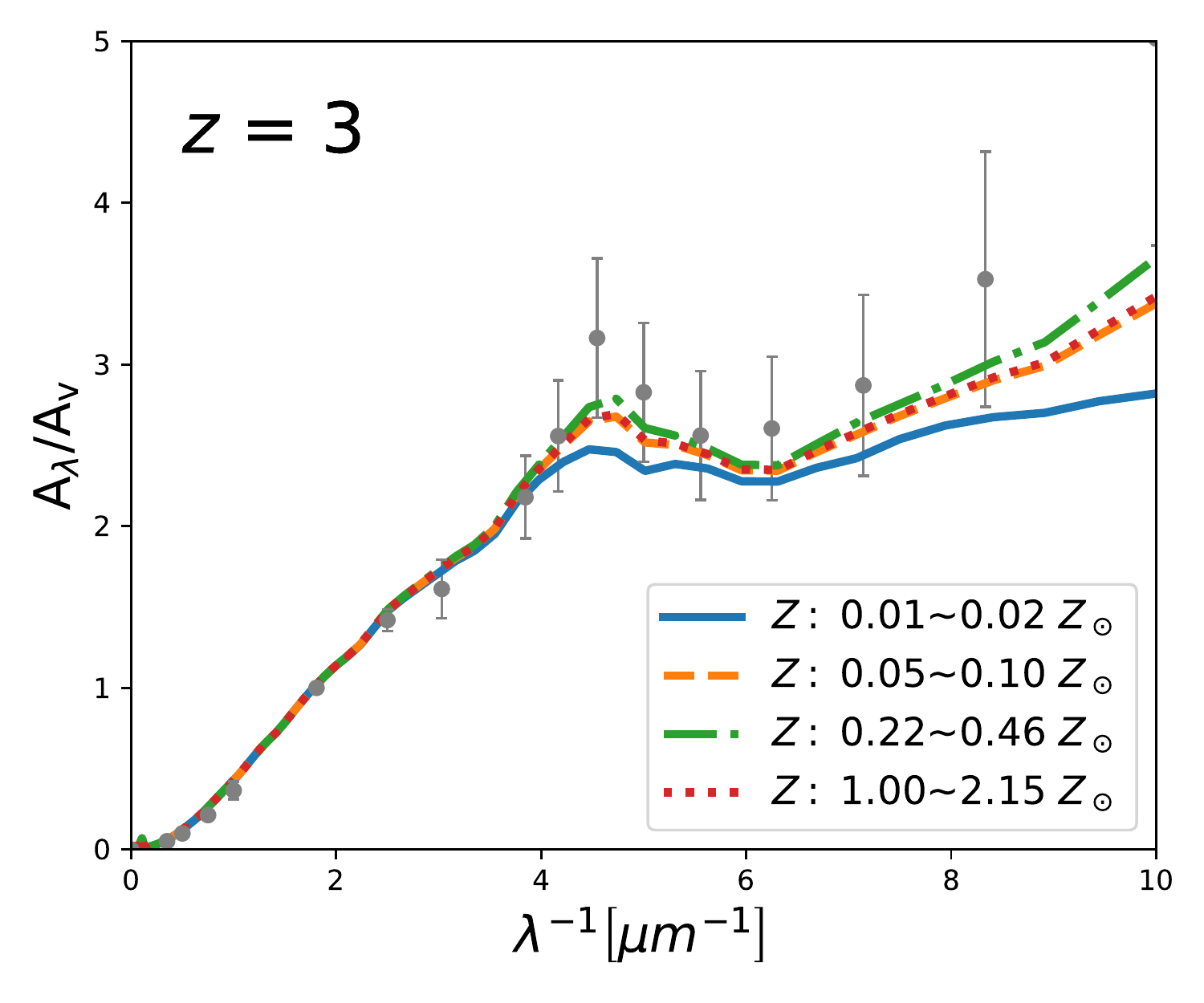}
	\includegraphics[width=0.32\textwidth]{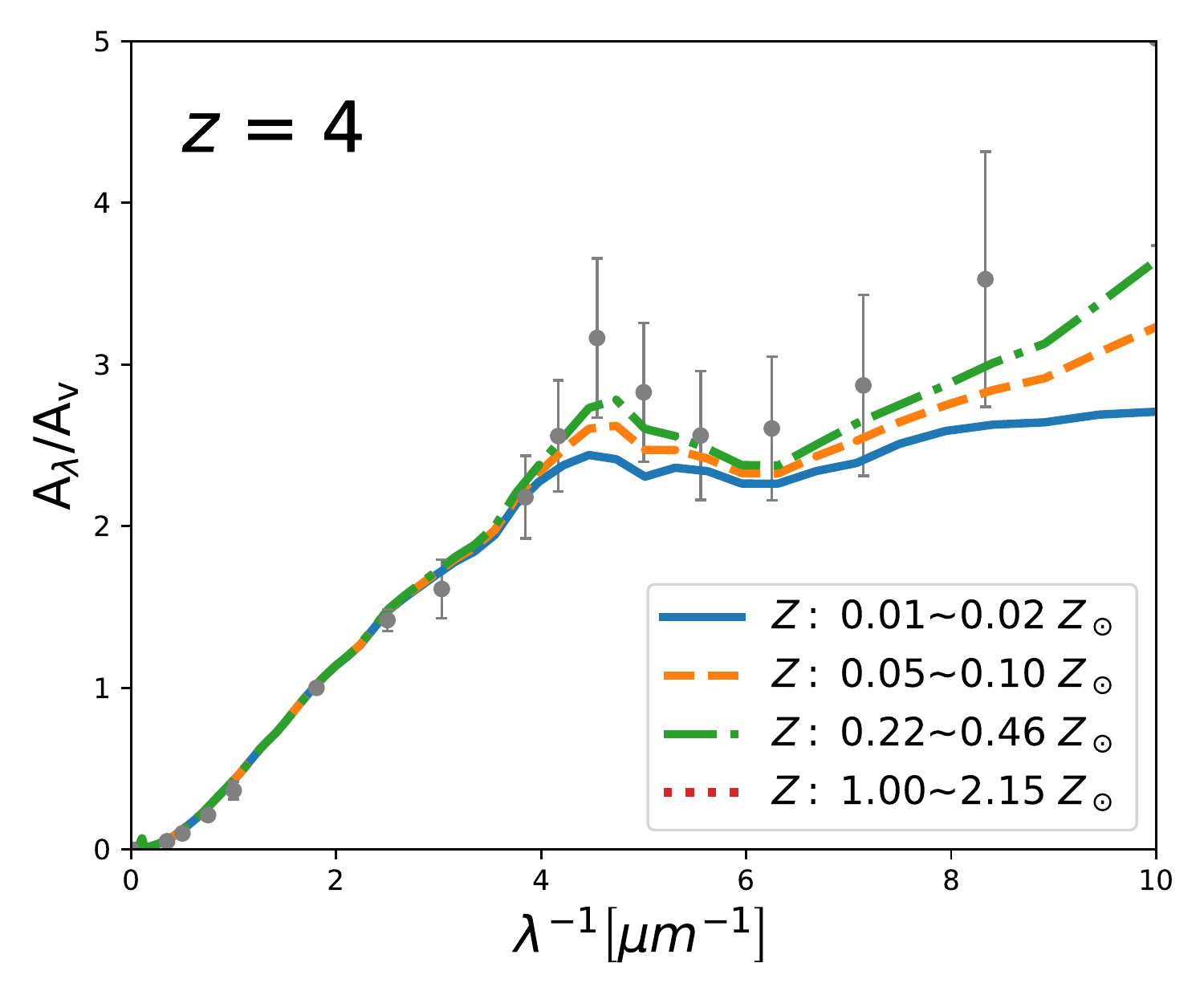}
	\includegraphics[width=0.32\textwidth]{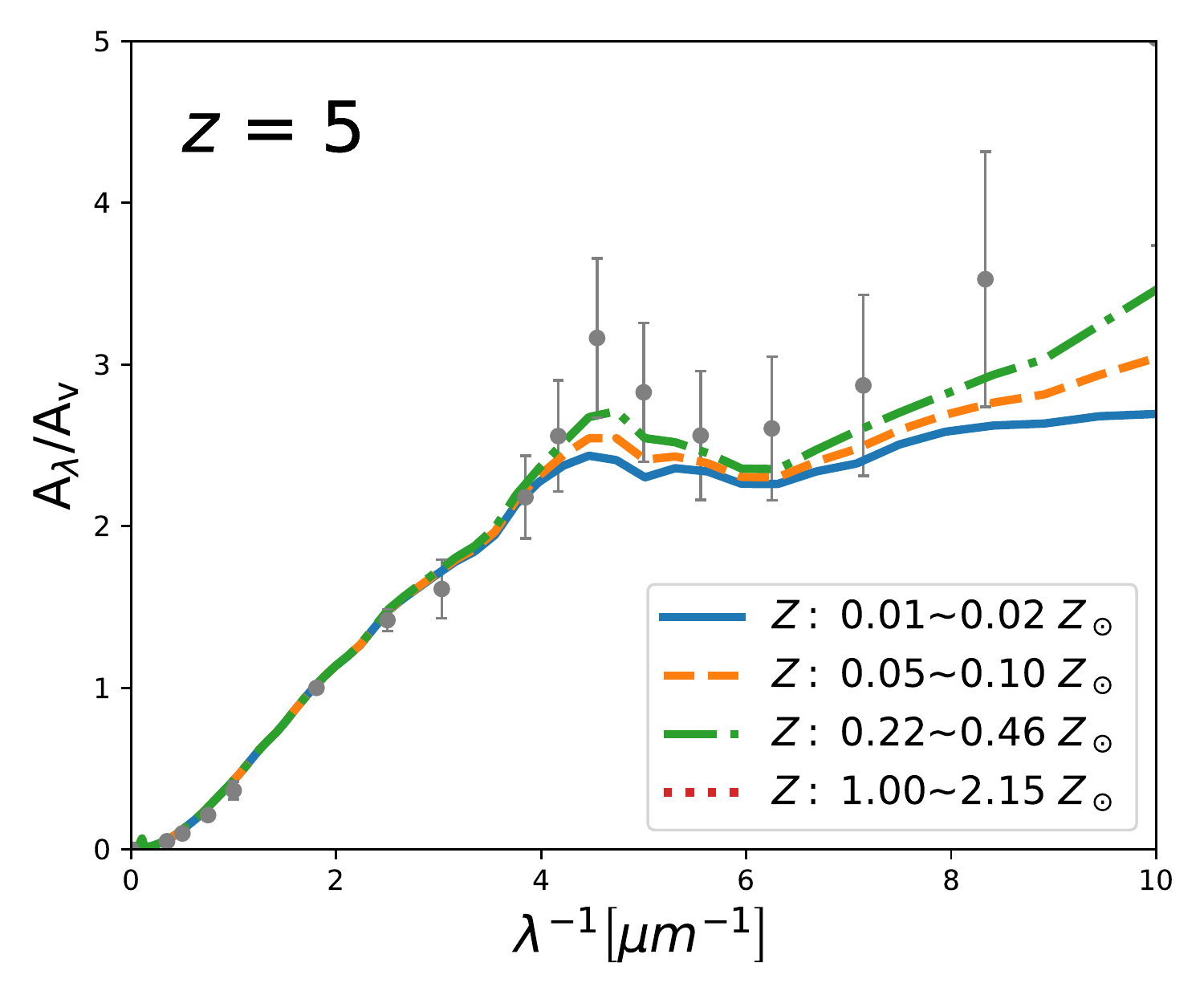}
	\caption{Redshift evolution of extinction curves from $z = 0$ to $5$.
	The redshift is indicated in each panel.
	We select four metallicity bins to present the metallicity dependence
	of extinction curves at each redshift as shown in the legend.
	The grey points with vertical error bars
	are the mean Milky Way extinction curve with 1$\sigma$ dispersion
	toward various lines of sight taken from \citet{Fitzpatrick:2007aa}.}
	\label{Fig:extinction_z}
	\end{center}
\end{figure*}

We examine the metallicity dependence of
extinction curves from $z = 5$ to 0 (Fig.~\ref{Fig:extinction_z}).
At $z \gtrsim 4$,
higher-metallicity galaxies
have steeper extinction curves because dust growth by accretion
increases the small-grain abundance.
The extinction curves become steeper from $z = 5$ to 3 in all metallicity
ranges, and the highest metallicity bin ($1 \lesssim Z \lesssim 2~Z_\odot$)
appears at $z = 3$ when the galaxies are sufficiently metal-enriched.
At $z\leq 4$, the steepest extinction curves are realised in galaxies with $Z \sim 0.3~ Z_\odot$
except at $z \sim 1$, where the steepest ones appear at a lower metallicity.
The steepest extinction curves are produced by the
effect of accretion, while the extinction curves become flatter at higher
metallicities because of coagulation. Thus, it is interesting to note that the
steepness of extinction curve is not a monotonic function of metallicity.
The redshift evolution is different in different metallicity ranges.
We find that the extinction curves for $Z \gtrsim 0.2\,Z_\odot$
become flatter from $z$= 3 to 1, which  is due to coagulation.
In contrast, the extinction curves with $Z \lesssim 0.2\,Z_\odot$
become steeper in the same redshift range because of accretion.
All extinction curves become steeper at $z = 0$,
because the diffuse gas hosting shattering is more prevalent in
galaxies at $z = 0$ than at $z \geq 1$.

Fig.~\ref{Fig:extinction_z} shows that the extinction curves at $z= 0$ are
basically in agreement with the Milky Way extinction curve estimated by
\citet{Fitzpatrick:2007aa} within 1$\sigma$ dispersion toward various lines of sight
\citep[see also][]{Nozawa:2013aa}.
This implies that we successfully included the relevant processes that
drive the dust evolution in galaxies.
Nevertheless, we note that
the extinction curves could be sensitive to the adopted
treatments of shattering and coagulation in the simulation;
in particular, we assumed that
shattering occurs in the diffuse medium while coagulation
in the dense medium with threshold density
$n_{\rm gas} = 0.1$ cm$^{-3}$ separating
the diffuse and dense medium.
Thus, there is still an uncertainty arising from the assumed value of the threshold.
However, we emphasize that the evolutionary trends discussed above
do not depend on the shattering and coagulation criteria.

Observationally, it is often easier to derive attenuation curves
if individual stars (or point sources) are not spatially resolved.
Attenuation curves include the radiation transfer effect inside galaxies
and usually differ from extinction curves
\citep[e.g.][]{Inoue:2005aa,Narayanan:2018aa}.
Therefore, for precise comparison with observed attenuation curves,
radiation transfer calculations are necessary.
Expecting that the relative steepness of attenuation curves still
reflects the relative abundance of small grains (at least statistically),
we attempt to compare our results with attenuation curves.
\citet{Kriek:2013aa} derived attenuation curves
for galaxies at 0.5 < $z$ < 2 from SED analyses.
They found that more galaxies with higher sSFR
have flatter attenuation curves and weaker bumps.
We also examined the sSFR dependence of extinction curves
in our simulation;
however, there is no clear dependence on sSFR because
the dispersion of $\stol$ is large at given sSFR
and the correlation between $\stol$ and sSFR is weak
as shown in Fig.\ \ref{Fig:S2L_1}c.
\citet{Salmon:2016aa} found that galaxies with larger colour excess
have a shallower attenuation curve slope using
a galaxy sample at $z \sim 1.5$--3. 
This could be interpreted as more efficient coagulation in more
dusty galaxies.
\citet{Cullen:2018aa} investigated star-forming galaxies
at $3 < z < 4$ 
and found that the attenuation curve shapes of their sample
are similar to the \citet{Calzetti:2000aa} law,
but not as steep as the Small Magellanic Cloud (SMC)
curve. They also obtained tentative evidence for steeper
attenuation curves at low stellar masses ($M_\ast\lesssim 10^9$ M$_{\sun}$),
while large grains dominate the dust abundance (thus, we expect flat
extinction curves) in low-mass galaxies in our model. We need to
include radiation transfer calculations to isolate the effects of extinction
curve shapes on the attenuation curves.

\citet{Cullen:2017aa}, using a cosmological simulation,
derived the attenuation curve slopes of galaxies at $z \sim 5$
that fit the luminosity function and the colour-magnitude relation.
They concluded that their results are consistent with the Calzetti curve
and that the SMC extinction law is ruled out in their simulation.
In our previous simulation of an isolated spiral galaxy \citep{Hou:2017aa},
we predicted spatially resolved extinction curves, and
reproduced the Milky Way extinction curve including its
dispersion in different lines of sight.
Since the internal dispersion of extinction curves is large,
more studies on the relation between extinction curves and
attenuation curves are required using higher resolution
cosmological simulations.

\section{discussion}
\label{section:discussion}

\subsection{Dust-to-gas ratio vs. Dust-to-stellar mass ratio}

Metallicity is usually used as an indicator of chemical enrichment.
In the same
way, dust-to-gas ratio is usually used to trace the dust enrichment
in a galaxy \citep{Lisenfeld:1998aa}.
Moreover, dust is usually coupled dynamically with the ISM on
a galactic scale. Thus, dust-to-gas ratio is a fair quantity that
traces the dust enrichment in the gas component of interest.
In contrast, dust-to-stellar mass ratio is affected by the dynamical
decoupling between dust (or gas) and stars.
Therefore, the comparison between observed and theoretical
dust-to-stellar mass ratios has uncertainties and complications in the spatial
extent of dust and stars. Indeed, dust emission is found to be
more extended than stellar emission \citep{Alton:1999aa}.
Moreover, a significant fraction of dust is suggested to be
contained in galaxy halos or in the CGM \citep{Menard:2010aa}
as theoretically confirmed in Paper I.
Because of such complication in
dust-to-stellar mass ratio, dust-to-gas ratio is
more preferable in testing dust enrichment models.

As shown in Section \ref{Result:D2G}, the relation between dust-to-gas ratio and metallicity clearly
shows the following important features in dust enrichment.
The relation reflects the dust condensation efficiency in stellar ejecta
at low metallicities. The non-linear relation appearing as a steep increase of
dust-to-gas ratio as a function of metallicity shows that dust
growth by accretion is the main dust producing mechanism.
Accretion is saturated at high metallicities,
where the dust-to-gas ratio is regulated by the balance
between accretion and SN destruction. Thus, analyzing the relation
between dust-to-gas ratio and metallicity gives us a clue to the efficiencies
of dust formation and destruction mechanisms.

From an observational point of view, however, stellar mass is easier to
derive compared with gas mass, because stellar emission usually gives
the first identification of distant galaxies using sensitive optical telescopes.
Radio observations of gas (H\,\textsc{i} and CO) emission are usually less
powerful in detecting a distant galaxy.
Thus, dust-to-stellar mass ratio is more convenient than dust-to-gas ratio
for distant galaxies.
We are able to extract roughly the same information on dust enrichment from
both dust-to-stellar mass ratio and dust-to-gas ratio, as we discussed in
Section~\ref{Result:D2S}.

Another advantage of dust-to-stellar mass ratio is that it
reveals the effect of astration. In contrast, dust-to-gas ratio is not affected by
astration, because both dust and gas are included into stars so that the
dust-to-gas ratio is unchanged. As argued in Section \ref{Result:D2S},
the decline of dust-to-stellar mass ratio at high metallicity is due to astration.
This also means that dust-to-stellar mass ratio is affected by how
efficiently the gas is converted to stars.


\subsection{Possible improvements}
\label{CompareObservation}

Our simulation seems to overproduce the dust abundance
around $M_\ast\sim 10^{10}$ M$_{\sun}$ as shown in
Figs.\ \ref{Fig:D2G}b and \ref{Fig:D2S}b. As argued in
Sections \ref{Result:D2G} and \ref{Result:D2S}, the discrepancy
cannot simply be attributed to
overproduction of dust and metals,
since we
succeed in reproducing the $\mathcal{D}$--$Z$ and $M_\ast$--$Z$
relations.

Some possible improvements on the theoretical side are worth discussing.
\citet{Hirashita:2017ab} presented a new model
of dust evolution with the AGN feedback cycle. They considered cyclic
gas cooling and heating, which lead to cyclic dust growth and destruction.
Therefore, AGN feedback could not
only heat the gas but also affect the dust evolution directly. This effect is not
fully taken into account in usual AGN feedback models. Although it is not
clear if this new feedback model resolves the above overproduction of dust,
it is worth implanting a new AGN feedback model in the future.

As shown in Section \ref{subsec:simulation_result} (Fig.\ \ref{Fig:Ms_Z}),
there may be a tendency that
the metallicities in low-$M_\ast$ ($M_\ast\lesssim 10^9$ M$_{\sun}$) galaxies
are underestimated in the simulation. As discussed there, we suspect that the lack of
spatial resolution leads to an underestimate of star formation activity and
chemical enrichment.
Therefore, if the star formation history is really affected by the spatial resolution
as indicated for low-mass galaxies in our simulation, a higher-resolution simulation
is desirable in the future to test the robustness of our prediction.

\subsection{Comparison with other theoretical studies}
\label{subsec:compare}

The dust mass function and various dust scaling relations
have been studied by cosmological simulations \citep{McKinnon:2017aa}
and semi-analytic models \citep{Popping:2017aa}.
\citet{Popping:2017aa} predicted that the dust mass function evolves little
from $z = 2$ to 0 at $M_\mathrm{d} \lesssim 10^{8.3}\,\Msun$,
whereas the number density of galaxies keeps increasing from $z = 2$ to 0
at higher dust masses.
Above $z = 2$, the galaxy number density decreases with increasing redshift
in the whole dust mass range.
\citet{McKinnon:2017aa} showed that
the galaxy number density decreases at $M_\mathrm{d} \lesssim 10^6\,\Msun$
from $z = 2.5$ to 0 and increases at $M_\mathrm{d} \gtrsim 10^6\,\Msun$.
Moreover, they did not predict enough dust-rich galaxies ($M_\mathrm{d} > 10^8\,\Msun$).
In our simulation, the change of the dust mass function is small
at low $M_\mathrm{d}$, which is qualitatively consistent with
\citet{Popping:2017aa}'s result. As pointed out above, \citet{McKinnon:2017aa}
assumed a higher dust yield by stars, which could be the reason why the
dust mass function increases at the low-mass end in their model.
At high $M_\mathrm{d}$, the dust mass function increases
from $z = 5$ to 1 and decreases from $z$ = 1 to 0 in our simulation.
This non-monotonic behaviour is different from both of the above simulations.
As discussed in Section \ref{RedshiftEvolution}, this non-monotonic behaviour
is a consequence of dust growth (accretion) and astration.
In our simulation, we included stronger dust growth than \citet{McKinnon:2017aa},
but such strong accretion is necessary
to explain the $\mathcal{D}$--$Z$ relation in our model. Therefore, it does not
seem that there is a perfect model that explain both dust mass function and
$\mathcal{D}$--$Z$ relation.

\citet{Popping:2017aa}'s semi-analytic model shows similar
$\mathcal{D}$--$Z$ relations at $z = 0$ to ours.
At $z > 0$, they indicate much weaker evolution of the $\mathcal{D}$--$Z$ relation
evolution than our results.
The different treatment of dust growth may be the reason for the difference.
Note that in our simulation, we solve the hydrodynamic evolution of gas in galaxies,
and the grain growth by accretion occurs in the dense gas in our simulated galaxies.
Therefore, accretion efficiency varies in a complex way.
On the other hand, in the semi-analytic model, instead of solving hydrodynamics,
they calculated the accretion time-scale by inferring the gas density in molecular clouds
from the surface density of SFR.

For the $M_{\rm d}/M_{\ast}$--$M_{\ast}$ relation,
\citet{McKinnon:2017aa} obtained a decreasing trend toward the high stellar-mass end,
and this trend does not evolve much from $z = 2.5$ to 0.
On the other hand, our $M_{\rm d}/M_{\ast}$--$M_{\ast}$ relation has
a peak at $M_{\ast} \sim 10^{10}\,\Msun$ at $z = 0$,
and the relation does evolve with redshift as shown in  Fig.~\ref{Fig:D2S}b
(because of the tight correlation between $M_\ast$ and $Z$, this figure
also shows the evolutionary trend in the $M_{\rm d}/M_{\ast}$--$M_{\ast}$ relation).
The complex behaviour of our result is due to the interplay between dust growth
and astration as mentioned above.
These diverse results could be tested by future observations.

\subsection{Prospect for the calculations of extinction curves}
\label{subsec:extinc}

We roughly reproduced the Milky Way extinction curve
at $z=0$ as shown in Fig.~\ref{Fig:extinction_z}, and confirmed that
the steepness of the FUV rise is a good indicator of the richness of
small grains relative to that of large grains.
It is well known that the SMC extinction curve has a steep
FUV rise but has a weak or even
no 2175\,\AA\ bump feature \citep[e.g.][]{Gordon:2003aa}.
\citet{Pei:1992aa} and \citet{Weingartner:2001aa} interpreted this as
deficiency of carbonaceous dust. On the other hand,
adopting amorphous carbon instead of graphite is also shown to be
a possible solution to eliminate the bump feature
\citep{Nozawa:2015aa,Hou:2016aa,Hou:2017aa}.
There are other types of carbonaceous dust such as hydrogenated
amorphous carbon \citep{Jones:2013aa};
however, our two-size approximation does not have enough capability
to investigate the detailed dust properties.
Therefore, we focus on the discussion about FUV slopes,
which is a more robust indicator of the small-to-large grain abundance ratio
compared with other features.

As shown in Section \ref{Result:extinction},
the steepest extinction curves are realized at
$Z \sim$ 0.3 $Z_\odot$, which corresponds to the highest
$\stol$ (Fig.~\ref{Fig:S2L_1}a).
This is consistent with the fact that the SMC and Large Magellanic Cloud (LMC)
have steeper extinction curves than the Milky Way;
the metallicities in the SMC and LMC are
$\sim$0.2 and $\sim$0.5~$Z_\odot$, respectively \citep[][]{Russell:1992aa}.
The SMC and LMC extinction curves are not reproduced successfully in our model
because of the prominent 2175 \AA\ bump,
although the FUV slope is steeper in galaxies with
$\sim 0.2-0.5\,Z_\odot$ than those with $1 Z_\odot$.
To reproduce the SMC extinction curve,
a smaller graphite-to-silicate mass ratio than that
of the Milky Way is necessary as proposed by \citet{Weingartner:2001aa}.
\citet{Bekki:2015ab}, based on an estimate of radiation pressure, suggested
that small carbonaceous dust grains can be removed selectively from galaxies
in starburst events. Under this assumption, they reproduced the SMC extinction curve.
\citet{Hou:2016aa} proposed a model in which small carbonaceous grains
are destroyed by SNe more efficiently than small silicate grains,
and produced a steep extinction curve similar to the SMC extinction curve.
In the future, it will be interesting to include the difference between these
species in cosmological simulations.


\subsection{Prospects for higher redshifts}
\label{subsec:highz}

As mentioned in Section \ref{RedshiftEvolution}, our simulation does not
have sufficient spatial resolution to predict meaningful statistical properties
at $z>5$.  To capture the star formation in first galaxies,
higher spatial and mass resolution is needed.
Nevertheless, some preliminary discussion is possible for
dust properties at $z>5$ based on our results.
At such high redshifts, the dust abundance is mostly dominated by
stellar dust production in most galaxies; thus, we expect that the dust-to-gas ratio
roughly follows the relation $\mathcal{D}=f_\mathrm{in}Z$.
Based on Fig. \ref{Fig:redshift},
the turning point shifts to higher metallicity with increasing $z$;
thus, we expect that only high-metallicity galaxies, which are too rare
to be sampled in our simulation box, can experience a drastic dust mass
increase by dust growth.
Indeed, the detection of dust becomes more and more difficult
if we go to higher redshift, especially at $z>5$, even by ALMA
\citep{Capak:2015aa,Bouwens:2016aa}, in spite of the
negative $K$-correction effect at submillimetre wavelengths.

There are a few examples of dust detections for `normal' galaxies
at $z$ > 6 \citep{Watson:2015aa,Willott:2015aa,Laporte:2017aa}.
To explain those dust-rich cases, \citet{Mancini:2015aa}
suggested an extremely efficient dust growth by accretion, which is
also supported by \citet{WangWC:2017aa}. Such extremely efficient growth
could be caused in very dense environments \citep{Kuo:2013aa},
which is difficult to realize in our simulation with a limited spatial resolution.
On the theoretical side, it is easier to focus on massive dusty galaxies,
although it requires a large simulation box size to obtain such rare objects.
\citet{Yajima:2015aa} performed a high-resolution zoom-in
cosmological simulation and focused on rare, heavily overdense regions
which can host high-$z$ quasars.
By assuming a constant dust-to-metal ratio,
they predicted a dust mass $M_{\rm d} \sim 4 \times 10^{10}$\,$\Msun$
in the most massive galaxy at $z \sim 6$, consistently with the current observations.

Based on Fig.\ \ref{Fig:redshift}c,
we expect that a galaxy with sub-solar metallicity is required to have
a high $\stol$ at higher redshift ($z > 5$).
We predict that those galaxies detected at $z\gtrsim 6$ by
ALMA have a high $\stol$ because dust growth by accretion not only
increases the total dust abundance but also raises the small grain abundance.
The extinction curves derived from high-$z$ quasars and gamma-ray bursts
provide opportunities to study the evolution of $\stol$.
At $z> 4$, the extinction curves tend to be flat
\citep{Maiolino:2004aa,Stratta:2007aa,Gallerani:2010aa},
which indicates that those sources have low small-to-large grain abundance ratios.
On the other hand, the extinction curves of lower-$z$
quasars show SMC-curve-like steepness \citep{Zafar:2015aa},
which implies a high $\stol$.
Our simulation generates a consistent evolutionary trend that
$\stol$ increases with decreasing redshift.


\section{Conclusions and Summary}
\label{section:conclusion}

To understand the dust evolution in galaxies statistically,
we perform a cosmological $N$-body/SPH simulation with implementation of
metal and dust enrichment. We treat dust evolution using the
two-size model \citep{Hirashita:2015aa},
which solves the production and destruction of large and small grains
in a consistent manner with the physical condition of gas.
This model represents the grain size distribution by the
abundances of small and large grains (separated at
grain radius $\sim 0.03~\micron$).
In the present work, we consider stellar dust production,
destruction in SN shocks and diffuse hot gas,
dust growth by accretion,
grain growth by coagulation and grain disruption by shattering.
While Paper I focused on the global properties of dust in the Universe,
this paper puts a particular emphasis on the basic scaling
relations of dust abundance indicators with main galaxy properties.
We newly implement a simple AGN feedback effect in this work.

We succeed in suppressing the dust mass function at
$M_\mathrm{dust} \gtrsim 10^9$\,M$_{\sun}$ by the AGN feedback.
Our simulation roughly reproduces the $z=0$ dust mass
function at $M_\mathrm{dust} \lesssim 10^8$\,M$_{\sun}$,
but there is still a significant excess at
$M_\mathrm{dust} \gtrsim 10^8\,\Msun$ (Fig.\,\ref{Fig:DustMF}).
The overproduction of galaxies with $M_\mathrm{dust} \gtrsim 10^8\,\Msun$
implies that the adopted AGN feedback model is too simple.

We examine various scaling relations
between dust properties (dust-to-gas ratio
and dust-to-stellar mass ratio) and characteristic
properties of galaxies such as metallicity ($Z$), stellar mass ($M_\ast$),
gas fraction ($f_\mathrm{gas}$) and sSFR.
Dust-to-gas ratio ($\mathcal{D}$) is a fundamental indicator of dust abundance.
We reproduce the observed $\mathcal{D}$--$Z$ relation at $z = 0$ (Fig.\,\ref{Fig:D2G}a).
This relation is interpreted as follows:
galaxies with $Z \lesssim 0.05~\mathrm{Z}_{\odot}$ basically follow the
relation ($\mathcal{D}=f_\mathrm{in}Z$) expected from the stellar dust production;
dust growth by accretion causes a steep increase of $\mathcal{D}$ at
$0.05 \lesssim Z \lesssim 0.5~\mathrm{Z}_{\odot}$;
$\mathcal{D}$ approaches $Z$ at $Z \gtrsim 0.5~\mathrm{Z}_{\odot}$
and dust growth by accretion is saturated.
A negative correlation between $\mathcal{D}$ and sSFR
is predicted, which is consistent with the observational data.
The $\mathcal{D}$--$f_{\rm gas}$ and $\mathcal{D}$--$M_\ast$
relations trace the $\mathcal{D}$--$Z$ relation because
$f_\mathrm{gas}$ and $M_\ast$ have strong negative and positive
correlations with $Z$, respectively.

From an observational point of view, it is easier to detect stellar emission
than gas emission, especially for distant galaxies. Therefore, we examine
another dust abundance indicator, dust-to-stellar mass ratio (Fig.\,\ref{Fig:D2S}).
At intermediate metallicities ($\sim 0.05 - 0.5\,\mathrm{Z}_\odot$),
$M_{\rm d}/M_{\ast}$ increases with increasing $Z$
because of dust growth by accretion.
At high metallicities ($Z \gtrsim 1\,Z_\odot$), in turn,
$M_{\rm d}/M_{\ast}$ decreases, because of astration.
The same trend occurs for the $M_{\rm d}/M_{\ast}$--$M_{\ast}$
and $M_{\rm d}/M_{\ast}$--$f_{\rm gas}$ relations:
$M_{\rm d}/M_{\ast}$ has a maximum around
$M_{\ast} \sim 10^{10}$\,$\Msun$
and $f_{\rm gas} \sim 0.4$.
$M_{\rm d}/M_{\ast}$ shows a weak positive correlation with sSFR
because galaxies with high $M_\mathrm{\ast}$ tend to have
experienced more astration.

Our simulation has an advantage of predicting
the small-to-large grain abundance ratio, $\stol$,
which represents the grain size distribution.
The ratio $\stol$ steeply increases with metallicity at $\lesssim 0.1\,Z_\odot$
because of accretion,
and it remains constant
at 0.1--0.3 Z$_\odot$ because large-grain formation by
coagulation is roughly in balance with small-grain production
by shattering and accretion (Fig.\,\ref{Fig:S2L_1}).
At $Z \gtrsim 0.3\,Z_\odot$,
$\stol$ decreases because of coagulation.
A similar trend is shown in the $\stol$--$M_\ast$ relation
since there is a tight correlation between metallicity and stellar mass.
The correlation between $\stol$ and sSFR is weak, and there is a weak trend
that galaxies with higher sSFR have lower $\stol$.
In the $\stol$--$f_{\rm gas}$ relation,
$\stol$ rises rapidly from $f_\mathrm{gas}\sim 1$ to $\sim 0.8$ and
it decreases toward the low-$f_{\rm gas}$ end.

To understand the redshift evolution of dust abundance from $z=0$ to 5,
we first examine the evolution of the $\mathcal{D}$--$Z$ relation (Fig.\,\ref{Fig:redshift}a).
The dust-to-gas ratio $\mathcal{D}$ is systematically higher as redshift decreases
at a fixed metallicity.
The metallicity at which dust growth by accretion overwhelms stellar
dust production (`the turning-point metallicity')
shifts to higher metallicity as the redshift increases.
Stronger SN destruction at higher redshift also creates a tendency of
lower $\mathcal{D}$ as the redshift increases.

We also investigate the evolution of the $M_{\rm d}/M_{\ast}$--$Z$ relation (Fig.\,\ref{Fig:redshift}b).
The peak of $M_{\rm d}/M_{\ast}$ shifts toward higher metallicity
as redshift decreases.
The maximum $M_{\rm d}/M_{\ast}$
increases from $z = 5$ to 1,
and decreases at $z \lesssim 1$.
This is caused by the complex balance between the increase of $\mathcal{D}$
by less destruction and
the decrease of $M_\mathrm{dust}$ by astration. 

We finally examine the redshift evolution of the $\stol$--$Z$ relation (Fig.\,\ref{Fig:redshift}c).
At $z \gtrsim 4$, a large fraction of low metallicity galaxies have low $\stol$
($<10^{-1.5}$).
After accretion become efficient, $\stol$ is raised to $\sim10^{-1}$.
Coagulation makes a decreasing trend of $\stol$ towards high metallicity.
$\stol$ at $z = 0$ is systematically higher than other redshift
because the fraction of diffuse gas, which hosts shattering, is higher.

Based on the obtained $\stol$,
we calculate extinction curves by adopting graphite and silicate for the
dust species.
The metallicity dependence of extinction curves are examined
from $z= 5$ to 0 (Fig.\,\ref{Fig:extinction_z}).
We find that galaxies with $Z \sim 0.3\,Z_{\odot}$ have the steepest
extinction curves in most of the redshift range,
and that the FUV slope of $Z \sim~ 0.1Z_{\odot}$
extinction curves steepens dramatically from $z= 5$ to 0.
Extinction curves with $Z \gtrsim 1\,Z_{\odot}$ become
shallower from $z= 3$ to 1 because of coagulation,
and extinction curves with $0.05\,\mathrm{Z}_{\odot} \lesssim Z \lesssim 0.5\,\mathrm{Z}_{\odot}$
become steeper at $z \lesssim 1$ because of shattering.
We successfully reproduce the Milky Way extinction curve at $z= 0$,
which implies that we have implemented the most relevant
dust evolution processes for Milky-Way-like galaxies.

Finally, we discuss the limitations of our simulation.
The following improvements will be worth trying in the future:
(\RNum{1}) The excess of massive galaxies in the stellar
and dust mass functions could be resolved
by including a more sophisticated treatment of AGN feedback.
(\RNum{2}) A higher spatial resolution will resolve more ISM structures
in low-mass galaxies and high-redshift galaxies,
and will give more accurate estimates of star formation and dust production there.
(\RNum{3}) Including variations of dust properties would predict a
greater variety in extinction curves as we observationally see in the difference
between the Milky Way and the SMC extinction curves.

\section*{Acknowledgements}

We thank Y.-H. Chu, T.-H. Chiueh, W.-H. Wang, T. Nozawa, and the anonymous referee
for useful comments.
HH is supported by the Ministry of Science and Technology
grant MOST 105-2112-M-001-027-MY3 and
MOST 107-2923-M-001-003-MY3 (RFBR 18-52-52-006).
KN and IS acknowledge the support from JSPS KAKENHI Grant Number JP17H01111.
KN acknowledges the travel support from the Kavli IPMU, World Premier Research Center Initiative (WPI), where part of this work was conducted.
KCH is supported by the IAEC-UPBC joint research foundation (grant No.\ 257)
and Israel Science Foundation (grant No.\ 1769/15).
We are grateful to V. Springel for providing us with the original version
of GADGET-3 code.
Numerical computations were carried out on Cray XC50 at the Center
for Computational Astrophysics,
National Astronomical Observatory of Japan and XL
at the Theoretical Institute for
Advanced Research in Astrophysics (TIARA) in Academia Sinica Institute of
Astronomy and Astrophysics (ASIAA).




\bibliographystyle{mnras}
\bibliography{kchou}


\bsp	
\label{lastpage}
\end{document}